\documentclass[aps,prd,twocolumn,superscriptaddress,amsmath,amssymb,nofootinbib,longbibliography,preprintnumbers,floatfix]{revtex4-2}
\usepackage[dvipsnames]{xcolor}
\usepackage[colorlinks]{hyperref}
\usepackage[normalem]{ulem}  
\usepackage{xspace}
\usepackage{bm}
\usepackage{comment}

\usepackage{todonotes}

\usepackage{tikz}
\usetikzlibrary{tikzmark}
\usetikzlibrary{positioning}
\usepackage[normalem]{ulem} 

\allowdisplaybreaks[1]

\hypersetup{linkcolor=BrickRed,citecolor=Green,
filecolor=Mulberry,
urlcolor=NavyBlue,
menucolor=BrickRed,
runcolor=Mulberry
}

\definecolor{lime}{HTML}{A6CE39}
\DeclareRobustCommand{\orcidicon}{\hspace{-1mm}
	\begin{tikzpicture}
	\draw[lime, fill=lime] (0,0) 
	circle [radius=0.16] 
	node[white] {{\fontfamily{qag}\selectfont \tiny \,ID}};
	\draw[white, fill=white] (-0.0525,0.095) 
	circle [radius=0.007];
	\end{tikzpicture}
	\hspace{-3mm}
}

\foreach \x in {A, ..., Z}{\expandafter\xdef\csname orcid\x\endcsname{\noexpand\href{https://orcid.org/\csname orcidauthor\x\endcsname}
			{\noexpand\orcidicon}}
}

\usepackage{graphicx}
\usepackage{bbm}

\makeatletter
\newcommand{\abs}{\@ifstar\abssmall\absbig}
\newcommand{\absbig}[1]{\left \lvert #1 \right \rvert}
\newcommand{\abssmall}[1]{\lvert #1 \rvert}
\makeatother

\renewcommand{\i}{\mathrm{i}}
\newcommand{\Tr}{\mathrm{Tr}}
\renewcommand{\Im}{\mathrm{Im}}

\newcommand{\dd}{\mathrm{d}}

\newcommand{\dv}[2]{\frac{\mathrm{d} #1}{\mathrm{d} #2}}

\newcommand{\bN}{\overline{N}}

\newcommand{\bP}{\overline{P}}

\newcommand{\ct}{c_{2 \theta}}
\newcommand{\st}{s_{2 \theta}}

\newcommand{\gNeq}{\mathfrak{g}^{(\mathrm{cl})}}
\newcommand{\bgNeq}{\bar{\mathfrak{g}}^{(\mathrm{cl})}}
\newcommand{\frg}{\mathfrak{g}}
\newcommand{\bfrg}{\bar{\mathfrak{g}}}
\newcommand{\Gammabar}{\overline{\Gamma}}
\newcommand{\bkappa}{\bar{\kappa}}
\newcommand{\asym}{{(\infty)}}

\begin{document}
\preprint{N3AS-25-008}

\title{Predicting the outcome of collisional neutrino flavor conversion}

\author{Julien Froustey\orcidA{}}
\email{jfroustey@berkeley.edu}
\affiliation{Department of Physics, University of California Berkeley, Berkeley, CA 94720, USA}
\affiliation{Department of Physics, University of California San Diego, La Jolla, CA 92093, USA}

\begin{abstract}
   Collisional flavor instabilities, driven by differing neutrino and antineutrino reaction rates, are expected to occur in dense astrophysical environments like supernovae and neutron star mergers, but have yet to be incorporated in large-scale simulations. We derive analytical expressions for the asymptotic state resulting from a homogeneous and isotropic instability, and apply these predictions to two representative conditions from a neutron star merger simulation. We emphasize the importance of using a collision term that allows for both damping of flavor coherence and relaxation back to the classical steady state. When this classical configuration is collisional-unstable, the resulting asymptotic state reflects a compromise between classical relaxation and flavor conversion, defining a “quantum” equilibrium with nonzero coherence. This analysis highlights the possibility of a tradeoff between classical and quantum effects, an important feature with regard to the inclusion of flavor oscillation physics into global simulations.
\end{abstract}

\maketitle

\section{Introduction}

In our current era of multimessenger astronomy, significant efforts are underway to model dense astrophysical objects such as core-collapse supernovae (CCSNe) or binary neutron star mergers (NSMs), with large-scale simulations incorporating a wide range of physics, including (magneto)hydrodynamics, general relativity, or the nuclear equation of state (see~\cite{Janka:2016fox,Burrows:2020qrp,Kyutoku:2021icp,Radice:2020ddv,Mezzacappa:2020pkk,Kiuchi:2024lpx} and references therein). A key ingredient in all simulations is neutrino transport~\cite{Mezzacappa:2020oyq,Foucart:2022bth,Fischer:2023ebq,Wang:2023tso} yet to date, neutrino oscillations are not systematically taken into account. Over the last decades, however, various flavor transformation mechanisms have been identified, which are expected to occur in these environments and may have important consequences on the dynamics (see, e.g.,~\cite{Duan:2010bg,Mirizzi:2015eza,Chakraborty:2016yeg,Tamborra:2020cul,Capozzi:2022slf,Volpe:2023met,Richers_review,Johns:2025mlm} for reviews). 

A direct approach, solving the “quantum” version of Boltzmann's equation describing neutrino transport with flavor mixing—the quantum kinetic equation (QKE)—is made unfeasible by the vast differences of length scales and timescales involved. Nevertheless, to assess the potential consequences of neutrino flavor oscillations, simplified models have been used in CCSNe~\cite{Stapleford:2019yqg,Ehring:2023lcd,Ehring:2023abs,Mori:2025cke}, postmerger accretion disks~\cite{Li:2021vqj,Just:2022flt,Fernandez:2022yyv}, and recently, a full NSM simulation~\cite{Qiu:2025kgy}. Other techniques, designed to mitigate the computational cost of a full multiangle QKE treatment, include an attenuation of the QKE Hamiltonian~\cite{Nagakura:2022kic,Xiong:2022vsy,Nagakura:2022xwe,Nagakura:2023mhr,Nagakura:2023wbf,Shalgar:2024gjt}, or the use of angular moments of the neutrino density matrix~\cite{Myers:2021hnp,Grohs:2022fyq,Grohs:2023pgq,Froustey:2023skf,Froustey:2024sgz,Kneller:2024buy,Grohs:2025ajr}. A convenient approach that can be embedded in classical simulations consists in designing a \emph{subgrid model} of flavor transformation~\cite{Nagakura:2023jfi}, which requires one to determine the state reached by the system post-flavor conversion. This has been heavily studied for fast flavor instabilities (FFIs), establishing that the final quasisteady state is characterized by some cancellation of the angular crossing between (anti)neutrino distributions, which can be predicted in various ways~\cite{Bhattacharyya:2020jpj,Bhattacharyya:2022eed,Zaizen:2022cik,Zaizen:2023ihz,Nagakura:2023xhc,Abbar:2023ltx,Xiong:2024pue,Richers:2024zit,George:2024zxz}. Such prescriptions have recently been used in global simulations of a postmerger disk~\cite{Lund:2025jjo} and CCSNe~\cite{Wang:2025nii}. 

A similar analytical prediction of the flavor conversion outcome remains, so far, unrealized for “collisional” flavor instabilities (CFIs), first evidenced in~\cite{Johns:2021qby}. CFIs take place as neutrino self-interactions combine with different neutrino/antineutrino reaction rates to amplify flavor coherence, contrary to the standard expectation that collisions damp flavor coherence. Linear stability analysis~\cite{Xiong:2022zqz,Lin:2022dek,Padilla-Gay:2022wck,Liu:2023pjw,Zaizen:2024faj} for a homogeneous and isotropic background has shown the existence of different collisional-unstable modes, with resonancelike and non-resonancelike regimes. The occurrence of CFIs has been investigated in actual environments, namely CCSNe~\cite{Xiong:2022vsy,Shalgar:2023aca,Liu:2023vtz,Akaho:2023brj,Liu:2024wzd,Xiong:2024tac} and NSMs~\cite{Xiong:2022zqz,Froustey:2023skf,Nagakura:2025hss}. However, most studies consider a simplified collision term \emph{which only acts on flavor coherence}. This still leads to a rich phenomenology, notably with multienergy neutrino gases, from flavor equipartition to flavor swaps~\cite{Lin:2022dek,Padilla-Gay:2022wck,Fiorillo:2023ajs,Johns:2023xae,Zaizen:2025ptx}. Including the flavor-diagonal parts of the collision term is nevertheless important, as they lead for instance to a flavor swap in the resonancelike case~\cite{Kato:2023cig}, instead of flavor equipartition.

This paper is structured as follows. In Sec.~\ref{sec:QKE}, we present our model of neutrino evolution, which includes both classical relaxation and coherence damping components of the collision term. We also present the associated linear stability results, with details in Appendix~\ref{app:LSA}. In Sec.~\ref{sec:evolution}, we numerically solve the evolution equations for configurations taken from a classical NSM simulation (see Appendix~\ref{app:attenuation} for a convergence study), and derive analytical predictions of the asymptotic state that agree with the nonlinear outcome of collisional flavor conversion. These results show significant differences with earlier studies that use different implementations of the collision term, which is examined in Sec.~\ref{sec:collision_term}. We summarize our findings and discuss the limitations and prospects of this study in Sec.~\ref{sec:outlook}.

Throughout this paper, we use natural units in which $\hbar = c = k_\mathrm{B} = 1$.

\section{Quantum Kinetic Equations} 
\label{sec:QKE}

\subsection{Model and assumptions}
\label{subsec:model}

For simplicity and in order to make a more direct connection with some global simulations that evolve three neutrino species ($\nu_e$, $\bar{\nu}_e$, and $\nu_x = \nu_\mu, \nu_\tau, \bar{\nu}_\mu, \bar{\nu}_\tau$) in energy-integrated schemes, we study a two-flavor, single-energy system. Focusing on situations where collisions play an important role on the dynamics, we also restrict ourselves to isotropic neutrino distributions. We thus seek to describe the evolution of the density matrix or, equivalently for isotropic distributions, the number density moment $N$. It is a $2 \times 2$ matrix in flavor space, where the diagonal elements ($N_{ee}$, $N_{xx}$) generalize classical number densities, and the off-diagonal component $N_{ex} \in \mathbb{C}$ encodes flavor coherence.

Furthermore, we consider for simplicity a homogeneous system, motivated by the results of~\cite{Liu:2023pjw} which show a relative dominance of the homogeneous mode of CFI. Under these assumptions, the evolution of $N$ is given by the QKEs~\cite{Sigl:1993ctk,Richers:2019grc}
\begin{equation}
\label{eq:QKE}
\begin{aligned}
    \dv{N}{t} &= - \i \left[H, N \right] + \frac12 \left\{ \begin{pmatrix} \kappa_e & 0 \\ 0 & \kappa_x \end{pmatrix} , N^\mathrm{(cl)} - N \right\} \, , \\
    \dv{\bN}{t} &= - \i \left[\overline{H}, \bN\right] + \frac12 \left\{ \begin{pmatrix} \bkappa_e & 0 \\ 0 & \bkappa_x \end{pmatrix} , \bN^\mathrm{(cl)} - \bN \right\}  \, .
\end{aligned}
\end{equation}
In this equation, the Hamiltonian $H$ is the sum of a vacuum, matter mean-field and self-interaction mean-field terms, namely, $H = H_\mathrm{vac} + H_\mathrm{mat} + H_{\nu \nu}$ for neutrinos and $\overline{H} = - H_\mathrm{vac} + H_\mathrm{mat} + H_{\nu \nu}$ for antineutrinos.\footnote{Some studies (e.g.,~\cite{Zhang:2013lka,Richers:2019grc,Grohs:2023pgq,Froustey:2023skf,Kato:2023cig,Kneller:2024buy,Grohs:2025ajr}) adopt a transposed convention for the antineutrino density matrix, which corresponds to the change $\bN \to \bN^*$.} 
The vacuum Hamiltonian reads in the flavor basis
\begin{equation}
    H_\mathrm{vac} = \frac{\omega}{2} \begin{pmatrix} - \ct & \st \\ \st & \ct \end{pmatrix} \, ,
\end{equation}
where $\omega \equiv \Delta m^2/2E_\nu$ with $E_\nu$ the common neutrino energy, and we use the compact notation $\ct = \cos(2 \theta)$ and $\st = \sin(2 \theta)$, with $\theta$ the mixing angle. In this work, we take $\Delta m^2 = 7.5 \times 10^{-5} \, \mathrm{eV^2}$ and $\sin^2(\theta) = 0.307$ (values corresponding to $\Delta m^2_{21}$ and $\theta_{12}$ in the 3-flavor case~\cite{ParticleDataGroup:2024cfk}). The matter term reads, also in the flavor basis,
\begin{equation}
    H_\mathrm{mat} = \sqrt{2} G_F \, n_e \begin{pmatrix} 1 & 0 \\ 0 & 0 \end{pmatrix} \equiv \lambda \begin{pmatrix} 1 & 0 \\ 0 & 0 \end{pmatrix} \, ,
\end{equation}
where $n_e$ is the difference of electron and positron number densities, which we express more conveniently as a function of the matter density ($\rho$) and the electron fraction ($Y_e$) through $n_e = Y_e \rho/m_u$, with $m_u \simeq 1.661 \times 10^{-24} \, \mathrm{g}$ the atomic mass unit. Finally, the self-interaction term is $H_{\nu \nu} = \sqrt{2} G_F (N - \bN)$. 

The last terms in Eq.~\eqref{eq:QKE} are an approximate description of collisions, using a “relaxation-time” approximation. The quantities $N^{(\mathrm{cl})}$, $\bN^{(\mathrm{cl})}$ are flavor-diagonal and describe the classical steady-state of the system. The timescales associated to this relaxation are given by the opacities $\kappa_e$, $\bkappa_e$ and $\kappa_x = \bkappa_x$ (we assume, consistently with many large-scale simulations, identical evolutions for $\nu_x$ and $\bar{\nu}_x$). For $\nu_e$ and $\bar{\nu}_e$, this corresponds to emission/absorption processes, with absorption opacity $\kappa_e$ and associated emissivity $\kappa_e N_{ee}^{(\mathrm{cl})}$ (likewise for $\bar{\nu}_e$), but the treatment of heavy-lepton flavor (anti)neutrinos is more subtle. Indeed, it was shown in~\cite{Liu:2024wzd} that a system in beta- and pair-equilibrium is stable to CFIs. In an actual system that is unstable to CFI, the heavy-lepton flavor neutrinos $\nu_x$ deviate from this thermal equilibrium as they are not perfectly trapped. The classical steady-state $N_{xx}^{(\mathrm{cl})}$ is thus the result of a competition between emission/absorption and advection. An effective homogeneous description of the system can be done in a relaxation-time approximation, assuming the same form of a collision-like term as for electron flavor (anti)neutrinos but with an effectively reduced emissivity (see Sec.~\ref{subsec:NSM} for an actual example), which leads to the form \eqref{eq:QKE} of the evolution equation. In the following, we conflate “collisions” with “classical steady-state relaxation,” bearing in mind that this remains a limitation of our homogeneous description. We note that this description would also apply to a perfectly homogeneous and isotropic system with a nonstandard classical equilibrium. Finally, since we consider isotropic neutrino distributions, the scattering processes can be discarded. 

We will assume that initially, the system is in the classical steady-state $N(t=0) = N^{(\mathrm{cl})} \equiv \mathrm{diag}\left(N_{ee}^{(\mathrm{cl})}, N_{xx}^{(\mathrm{cl})}\right)$, and likewise for antineutrinos. This assumption allows us to focus on the competing effects of flavor conversion and classical relaxation. If the initial configuration was not a classical steady-state, the diagonal components would evolve on the collisional timescale while flavor coherence grows, but accounting for these effects would notably require to describe the advection of $\nu_x$. We thus focus on an idealized situation that allows for an analytical treatment, and provides guidance for future studies on larger spatial scales.

\subsection{Linear regime}

We first establish the main features of the CFI in the linear regime, where $\lvert N_{ex} \rvert \ll 1$. Following~\cite{Lin:2022dek}, we write the $ex$ components of the QKEs as
\begin{equation}
\label{eq:QKE_linear}
    \i \frac{\dd}{\dd t}\begin{bmatrix} N_{ex} \\ \bN_{ex} \end{bmatrix} \simeq \frac{\omega \st}{2 \sqrt{2} G_F} \begin{bmatrix} - \mathfrak{g} \\ \bar{\mathfrak{g}} \end{bmatrix} + \Lambda  \begin{bmatrix} N_{ex} \\ \bN_{ex} \end{bmatrix} \, ,
\end{equation}
where
\begin{equation}
    \Lambda = \begin{bmatrix} - \omega \ct + \lambda - \bar{\mathfrak{g}} - \i \Gamma & \mathfrak{g} \\
    - \bar{\mathfrak{g}} & \omega \ct + \lambda + \mathfrak{g} - \i \overline{\Gamma} \end{bmatrix} \, ,
\end{equation}
with the compact notations $\mathfrak{g} \equiv \sqrt{2} G_F (N_{ee} - N_{xx})$ and $\bar{\mathfrak{g}} \equiv \sqrt{2} G_F (\bN_{ee} - \bN_{xx})$, and the average collision rates $\Gamma \equiv (\kappa_e + \kappa_x)/2$ and $\Gammabar \equiv (\bkappa_e + \bkappa_x)/2$. The solution of Eq.~\eqref{eq:QKE_linear} is
\begin{equation}
\label{eq:Nex_sol}
    \begin{bmatrix} N_{ex}(t) \\ \bN_{ex}(t) \end{bmatrix} = Q_0 + Q_+ e^{- \i \Omega_+ t} + Q_- e^{-\i \Omega_- t} \, ,
\end{equation}
where $Q_{\pm}$ are the “plus” and “minus” eigenmodes of $\Lambda$, associated with the eigenvalues $\Omega_\pm$, and $Q_0$ is set by the initial condition $N_{ex}(t=0) = \bN_{ex}(t=0)=0$. The imaginary parts of $\Omega_\pm$, which correspond to the growth rate of the instability if they are positive, and associated eigenvectors are (see Appendix~\ref{app:LSA} for details)
\begin{subequations}
\label{eq:LSA_CFI}
\begin{align}
\label{eq:CFI_minus}
    \Im\left(\Omega_-\right) &= \frac{\Gammabar \bfrg - \Gamma \frg}{\frg - \bfrg} \, ,  & Q_- &\propto \begin{bmatrix} \frg \\\bfrg \end{bmatrix} \, , \\
    \Im\left(\Omega_+\right) &= \frac{\Gamma \bfrg - \Gammabar \frg}{\frg - \bfrg} \, , & Q_+ &\propto \begin{bmatrix} 1 \\ 1 \end{bmatrix} \, . \label{eq:CFI_plus}
\end{align}
\end{subequations}
We assumed that we are outside the resonancelike regime, which is the case in the examples studied below. The above derivation is valid for the initial configuration where $N_{ex} \simeq 0$ and $N_{aa} = N_{aa}^{(\mathrm{cl})}$. Specifically, Eq.~\eqref{eq:LSA_CFI} with $\frg = \gNeq$ indicates whether this initial configuration is (un)stable. More generally, the linear stability results \eqref{eq:LSA_CFI} are applicable at all times where $|N_{ex}(t)| \ll 1$, where they give in particular the instantaneous instability growth rate.

\section{Evolution and asymptotic state}
\label{sec:evolution}

\subsection{NSM examples} 
\label{subsec:NSM}

\begin{figure}[!ht]
    \centering
    \includegraphics[width=\columnwidth]{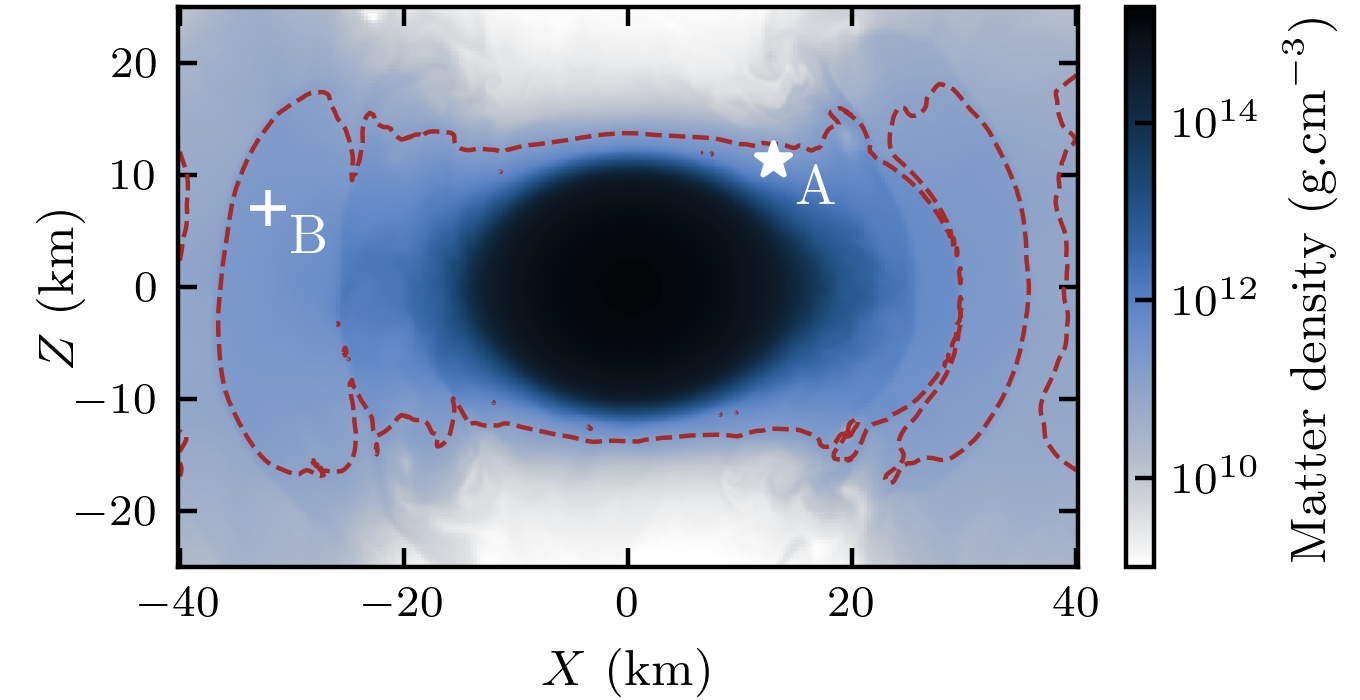}
    \caption{Poloidal slice from the 7 ms postmerger snapshot of the “M1-NuLib” NSM simulation in~\cite{Foucart:2024npn}. The dashed line is the contour of $\nu_e$ flux factor equal to $0.1$.}
    \label{fig:NSM_snapshot}
\end{figure}

We consider two configurations obtained from a general relativistic simulation of the merger of two neutron stars with component masses of $1.3 \, M_\odot$ and $1.4 \, M_\odot$~\cite{Foucart:2024npn}, specifically, from the 7 ms postmerger snapshot shown on Fig.~\ref{fig:NSM_snapshot} (see~\cite{Richers:2024zit} for a discussion of FFIs in this snapshot). These points (“Point A” and “Point B”) are chosen to illustrate two cases where a different branch of the CFI is unstable, namely, the minus (plus) mode for Point A (B). We detail the relevant parameters in Table~\ref{tab:NSMpoints}. The opacities appearing in Eq.~\eqref{eq:QKE} are the absorption opacities obtained from NuLib~\cite{OConnor:2014sgn} assuming the SFHo equation of state (like in the NSM simulation the data is extracted from). As discussed before, we assume that the values $N_{aa}$, $\bN_{aa}$ in Table~\ref{tab:NSMpoints} correspond to the classical steady-state\footnote{This classical state is not a genuine chemical equilibrium (for which $N_{xx}$ would be between $N_{ee}$ and $\bN_{ee}$), but we effectively describe the escape of $\nu_x$ through the value $N_{xx}^{(\mathrm{cl})} < N_{ee}^{(\mathrm{cl})},\bN_{ee}^{(\mathrm{cl})}$.} $N_{aa}^{(\mathrm{cl})}$, $\bN_{aa}^{(\mathrm{cl})}$, which is our initial condition.

\begin{table}[!ht]
    \centering
    \setlength{\tabcolsep}{8pt}
    \begin{tabular}{|r|c|c|}
    \hline
        & Point A & Point B \\ \hline \hline
        $N_{ee} \ (\mathrm{cm}^{-3})$ & $1.04 \times 10^{34}$ &  $5.10 \times 10^{33}$ \\
        $\bN_{ee} \ (\mathrm{cm}^{-3})$ & $1.26 \times 10^{34}$ &  $4.49 \times 10^{33}$ \\
        $N_{xx} = \bN_{xx} \ (\mathrm{cm}^{-3})$ & $7.10 \times 10^{33}$ &  $1.23 \times 10^{33}$ \\ \hline
        $\rho \ (\mathrm{g.cm^{-3}})$ & $2.92 \times 10^{11}$ & $2.50 \times 10^{11}$ \\
        $Y_e$ & $0.298$ & $0.231$ \\
        $\langle E_\nu \rangle \ (\mathrm{MeV})$ & $35.7$ & $34.6$ \\ \hline
        $\kappa_e \ (\mathrm{s}^{-1})$ & $7.63 \times 10^5$ & $2.78 \times 10^5$ \\
        $\bkappa_e \ (\mathrm{s}^{-1})$ & $2.11 \times 10^5$ & $5.95 \times 10^4$ \\
        $\kappa_x = \bkappa_x \ (\mathrm{s}^{-1})$ & $2.87 \times 10^3$ & $5.82 \times 10^2$ \\ \hline
    \end{tabular}
    \caption{Conditions at points A and B identified on Fig.~\ref{fig:NSM_snapshot}. The flux factors are smaller than $0.1$ for all species and are neglected in this study.
    }
    \label{tab:NSMpoints}
\end{table}

\subsection{Numerical results}
\label{subsec:numerical}

\begin{figure*}[!ht]
    \centering
    \includegraphics[width=\columnwidth]{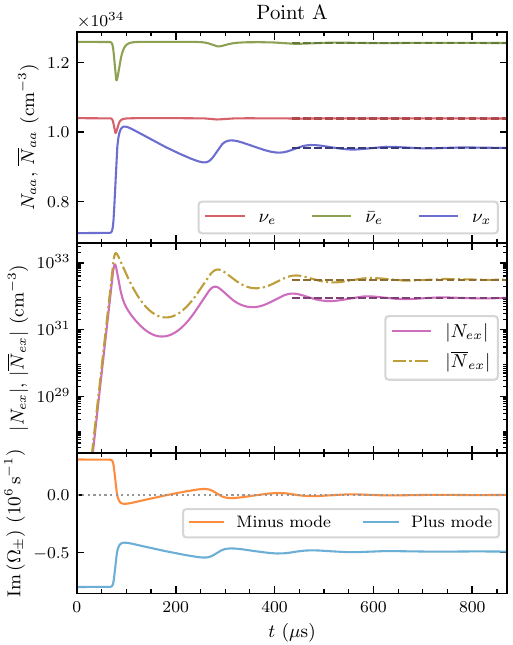}
    \includegraphics[width=\columnwidth]{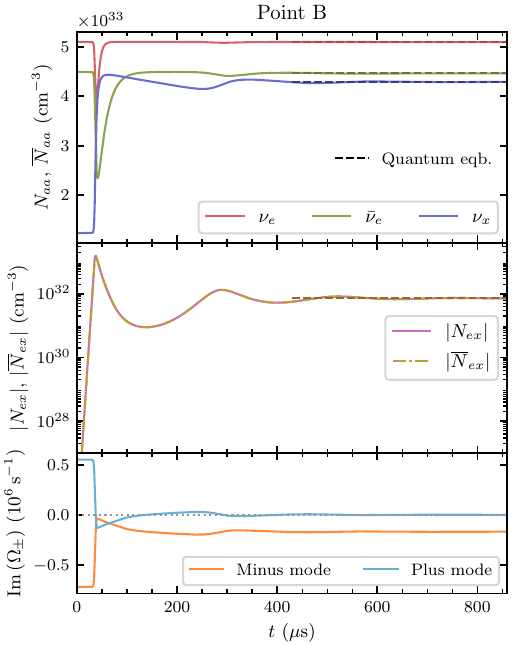}
    \caption{Evolution of the number density flavor on-diagonal (top panels) and off-diagonal (middle panels) components for Point A (left) and Point B (right). The asymptotic state predictions (see text) are shown in dashed lines. The bottom panels show the instantaneous growth/decay rate of each instability mode, as given by Eq.~\eqref{eq:LSA_CFI} with $\mathfrak{g}(t)$ and $\bar{\mathfrak{g}}(t)$.}
    \label{fig:pointsAB}
\end{figure*}

We solve the QKEs~\eqref{eq:QKE} with the explicit Runge-Kutta method of order 5(4) implemented in the function \texttt{solve\_ivp} of SciPy~\cite{2020SciPy-NMeth}. Because of the large difference of timescales involved, we avoid issues of numerical convergence by multiplying the Hamiltonian $H$ by an attenuation factor $\eta = 10^{-3}$, but we show in Appendix~\ref{app:attenuation} that the results are unaffected as long as $\eta > 10^{-4}$ (see Fig.~\ref{fig:attenuation_factor}). We plot the evolution of $N$ and $\bN$ on Fig.~\ref{fig:pointsAB} for Point A (left) and B (right), along with the evolution of the growth rate of each mode, as obtained from Eq.~\eqref{eq:LSA_CFI}, at each time step of the calculation. The maximum value of $|N_{ex}|/\Tr[N]$ is 0.05 (0.2) for Point A (B), which justifies to use these linear stability analysis formulas to assess the stability of each instantaneous configuration. We note that the vacuum term, which is subdominant, seeds the instability but plays no significant role otherwise, such that $N_{xx} \simeq \bN_{xx}$ at all times.

For Point A, the first $100 \, \mathrm{\mu s}$ correspond to the growth of the unstable minus mode (exponential growth of $|N_{ex}|$ and $|\bN_{ex}|$, see middle panel), which leads to flavor conversion $\nu_e, \bar{\nu}_e \to \nu_x$. After the instability saturates, the densities of $\nu_e$ and $\bar{\nu}_e$ are repopulated very close to their classical equilibrium values, and oscillations have resulted in a net increase of $N_{xx}$. At the end of this first phase, $N_{xx} \simeq N_{ee}$. As can be seen in the bottom panel, this configuration is stable, consistent with Eq.~\eqref{eq:CFI_minus} with $\mathfrak{g} \simeq 0$. However, on a timescale $\propto \kappa_x^{-1}$, $N_{xx}$ is drawn back toward $N_{xx}^{(\mathrm{cl})}$, which brings the system into the unstable domain; the off-diagonal components grow again, and subsequent flavor conversion occurs, albeit with a smaller amplitude. After several cycles, the system converges toward a “quantum” equilibrium, where in particular $N_{xx} \neq N_{xx}^{(\mathrm{cl})}$ and flavor coherence is nonzero (with $|N_{ex}|/\Tr(N)$ at the percent level). For Point B, the conclusions are identical, reversing the roles of $\nu_e$ and $\bar{\nu}_e$, and the minus/plus modes. The difference of instability branches between A and B is visible in the evolutions of $N_{ex}$ and $\bN_{ex}$, whose different behavior is consistent with the difference between $Q_-$ and $Q_+$ [see Eq.~\eqref{eq:LSA_CFI}].

\subsection{Prediction of the asymptotic state}
\label{subsec:predict_asym}

Our goal is to analytically predict the values taken by $N$ and $\bN$ for $t \to \infty$, which we will denote with a superscript $(\infty)$. More specifically, in the spirit of designing a semiclassical subgrid model which provides a prescription for the evolution of the classical neutrino distributions (so without treating flavor coherence, even though it \emph{is} crucially nonzero in the final state), we want to predict the asymptotic values of the flavor-diagonal components $N_{aa}$ and $\bN_{aa}$. As indicated by the numerical results of Fig.~\ref{fig:pointsAB}, this asymptotic state is characterized by \emph{(i)} constant flavor-diagonal number densities, and \emph{(ii)} a configuration lying \emph{at the edge of collisional instability}. Interestingly, a similar “edge of instability” configuration was obtained in~\cite{Fiorillo:2024qbl} for a system externally driven to a (fast) instability, and is reminiscent of more general considerations on the equilibrium of a neutrino system, as seen in the framework of plasmonlike collective flavor conversions (see, e.g.,~\cite{Fiorillo:2024uki,Fiorillo:2024pns,Fiorillo:2025npi}) or miscidynamics~\cite{Johns:2023jjt,Johns:2023xae}.

As stated above, the vacuum term contribution is negligible for the nonlinear evolution, such that $N_{xx} \simeq \bN_{xx}$ at all times. Steady-state equations for condition \emph{(i)} can be written as $\mathrm{Tr}(\dot{N}) = 0$ and $\dot{N}_{ee} - \dot{\bN}_{ee} = 0$, that is
\begin{equation}
\label{eq:asym_1}
    \begin{aligned}
        \kappa_e[{N}_{ee}^\asym - N_{ee}^{(\mathrm{cl})}] &= - \kappa_x[{N}_{xx}^\asym - N_{xx}^{(\mathrm{cl})}] \, , \\
        \kappa_e[{N}_{ee}^\asym - N_{ee}^{(\mathrm{cl})}] &= \bkappa_e[{\bN}_{ee}^\asym - \bN_{ee}^{(\mathrm{cl})}] \, .
    \end{aligned}
\end{equation}
Condition \emph{(ii)} reads differently depending on which mode was unstable. If the system was in a “minus” instability, the final state satisfies $\Im(\Omega_-^{(\infty)}) = 0$, that is
\begin{equation}
\label{eq:edge_instab_minus}
    \Gamma [{N}_{ee}^\asym - {N}_{xx}^\asym] = \Gammabar [{\bN}_{ee}^\asym - {N}_{xx}^\asym] \, ,
\end{equation}
while a system entering a “plus” instability reaches the final state determined by $\Im(\Omega_+^{(\infty)})=0$, i.e.,
\begin{equation}
\label{eq:edge_instab_plus}
     \Gamma [{\bN}_{ee}^\asym - {N}_{xx}^\asym] = \Gammabar [{N}_{ee}^\asym - {N}_{xx}^\asym] \, .
\end{equation}
Combining the previous equations leads to the expression for the asymptotic value of $N_{ee}$,
\begin{equation}
\label{eq:Neeasym_minus}
    \sqrt{2}G_F [{N}_{ee}^\asym - N_{ee}^\mathrm{(cl)}] = \frac{1}{\kappa_e} \frac{\Gammabar \bgNeq - \Gamma \gNeq}{\Gamma \left(\frac{1}{\kappa_e}+\frac{1}{\kappa_x}\right) - \Gammabar \left(\frac{1}{\bkappa_e}+\frac{1}{\kappa_x}\right)}
\end{equation}
for the minus instability, and
\begin{equation}
\label{eq:Neeasym_plus}
    \sqrt{2} G_F[{N}_{ee}^\asym - N_{ee}^\mathrm{(cl)}] = \frac{1}{\kappa_e} \frac{\Gamma \bgNeq - \Gammabar \gNeq}{\Gammabar \left(\frac{1}{\kappa_e}+\frac{1}{\kappa_x}\right) - \Gamma \left(\frac{1}{\bkappa_e}+\frac{1}{\kappa_x}\right)}
\end{equation}
for the plus instability. Equation~\eqref{eq:Neeasym_minus} or \eqref{eq:Neeasym_plus}, along with Eq.~\eqref{eq:asym_1}, fully determine the asymptotic state of the flavor-diagonal components of the density matrix. These predictions are shown as dashed lines on Fig.~\ref{fig:pointsAB}, with a perfect agreement with the numerical solution. Taking into account the CFI, the classical steady-state $N^{(\mathrm{cl})}$ is replaced by the quantum one $N^{(\infty)}$.

\paragraph*{Generalizations —} Recent simulations (see e.g.,~\cite{Bollig:2017lki,Fischer:2020vie,Loffredo:2022prq,Ng:2024zve,Pajkos:2025oyf}) have shown that muons could be present and have a significant impact on the dynamics in CCSNe or NSMs. For such situations, we could not assume $N_{xx} \simeq \bN_{xx}$ and would have to consider a full 4-species treatment. The derivation of the asymptotic state is straightforward to generalize, with $[\bN_{ee}^\asym - N_{xx}^\asym] \to [\bN_{ee}^\asym - \bN_{xx}^\asym]$ in Eqs.~\eqref{eq:edge_instab_minus}--\eqref{eq:edge_instab_plus} and with the additional (now independent) equation $\Tr(\dot{\bN})=0$, reading $\bkappa_e [\bN_{ee}^\asym - \bN_{ee}^{(\mathrm{cl})}]=-\bkappa_{x}[\bN_{xx}^\asym - \bN_{xx}^{(\mathrm{cl})}]$ (note that now $\bkappa_x \neq \kappa_x$). Although such a configuration is outside the scope of this paper and has not been numerically studied, we postulate that these small changes would describe the CFI asymptotic state with $\nu_x \neq \bar{\nu}_x$.

Strictly speaking, four species would not be enough and one should consider in general configurations where $\nu_\mu \neq \bar{\nu}_\mu \neq \nu_\tau \neq \bar{\nu}_\tau$, which we leave for future work. We note that the two-flavor scheme developed here could still be useful to determine a three-flavor asymptotic state, similarly to the method employed in~\cite{Liu:2025tnf} in the fast regime.

\subsection{Magnitude of flavor coherence} 

\begin{figure}[!ht]
    \centering
    \includegraphics[width=0.98\columnwidth]{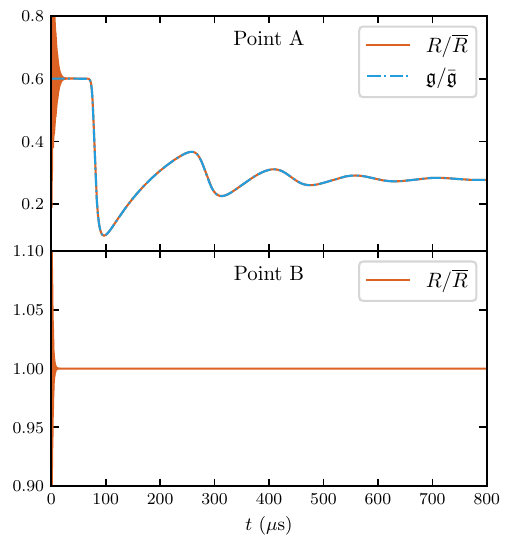}
    \caption{Conservation of the unstable eigenvector structure in flavor coherence for the two points studied. \emph{Top:} Point A, where the minus mode is unstable. \emph{Bottom:} Point B, where the plus mode is unstable. The ratio $R/\overline{R}$ follows the ratio of components of $Q_\pm$, see Eq.~\eqref{eq:LSA_CFI}.}
    \label{fig:eigvect_t}
\end{figure}

In the previous section, we focused on the quantum steady-state reached by the flavor on-diagonal components of $N$. We now provide a more complete picture of this asymptotic state, discussing flavor coherence. To that end, we introduce the modulus and phase of the flavor off-diagonal component of the density matrix, specifically,
\begin{equation}
    \begin{aligned}
        N_{ex} &\equiv R \, e^{i S} \, , &&\ R>0 \, , \ 0 \leq S < 2 \pi \, , \\
        \bN_{ex} &\equiv \overline{R} \, e^{i \overline{S}} \, , &&\ \overline{R}>0 \, , \ 0 \leq \overline{S} < 2 \pi \, ,
    \end{aligned}
\end{equation}
and we define the phase difference $\Delta S \equiv \overline{S} - S$. The evolution of $N_{ex}$ and $\bN_{ex}$ can thus be written
\begin{subequations}
\begin{align}
    \dot{R} &= - \Gamma \, R + \mathfrak{g} \, \overline{R} \, \sin(\Delta S) \, , \\
    R \, \dot{S} &= (\omega \, \ct -\lambda) R - \mathfrak{g} \, \overline{R} \, \cos(\Delta S) + \bar{\mathfrak{g}} \, R \, , \\ 
    \dot{\overline{R}} &= - \Gammabar \, \overline{R} + \bar{\mathfrak{g}} \, R \, \sin(\Delta S) \, , \\
    \overline{R} \, \dot{\overline{S}} &= (- \omega \, \ct - \lambda) \overline{R} + \bar{\mathfrak{g}} \, R \, \cos(\Delta S) - \mathfrak{g} \, \overline{R} \, .
\end{align}
\end{subequations}
The evolution of $N_{ee}$ reads, in terms of moduli and phases of the off-diagonal components and neglecting the vacuum term,
\begin{equation}
    \dot{N}_{ee} = - 2 \sqrt{2} G_F \,  R \, \overline{R} \, \sin(\Delta S) - \kappa_e[N_{ee} - N_{ee}^{(\mathrm{cl})}] \, .
\end{equation}
In Sec.~\ref{subsec:predict_asym}, we have defined the post-CFI asymptotic state based on the constancy of the flavor-diagonal components, and a “threshold of instability” condition. We can revisit this latter feature based on the following observations. When considering the off-diagonal components, the asymptotic state is further characterized by the constancy of the magnitudes $R$, $\overline{R}$ (see middle panels of Fig.~\ref{fig:pointsAB}). In addition, the ratio $R/\overline{R}$ remains determined by the unstable eigenvector $Q_\pm$ throughout the evolution, which is confirmed in Fig.~\ref{fig:eigvect_t} for both Points A [top panel, ratio $R/\overline{R}$ equal at all times to $\mathfrak{g}/\bar{\mathfrak{g}}$, consistent with Eq.~\eqref{eq:CFI_minus}] and B [bottom panel, ratio $R/\overline{R} = 1$, consistent with Eq.~\eqref{eq:CFI_plus}]. At small times, the large scatter of $R/\overline{R}$ corresponds to the system settling into the unstable mode. 

The asymptotic state thus satisfies the following set of equations:
\begin{subequations}
\begin{align}
    &\kappa_e[{N}_{ee}^\asym - N_{ee}^{(\mathrm{cl})}] = - \kappa_x[{N}_{xx}^\asym - N_{xx}^{(\mathrm{cl})}] \, , \\
    &\kappa_e[{N}_{ee}^\asym - N_{ee}^{(\mathrm{cl})}] = \bkappa_e[{\bN}^\asym_{ee} - \bN_{ee}^{(\mathrm{cl})}] \, , \\
    &\kappa_e[{N}^\asym_{ee} - N_{ee}^{(\mathrm{cl})}] = - 2 \sqrt{2} G_F \, {R}^\asym \, {\overline{R}}^\asym \, \sin[\Delta {S}^\asym] \, , \label{eq:fullasym_Nee} \\
    &\Gamma \, {R}^\asym = \sqrt{2} G_F [{N}^\asym_{ee} - {N}^\asym_{xx}] {\overline{R}}^\asym \sin[\Delta {S}^\asym] \, , \label{eq:fullasym_R} \\
    &\Gammabar \, \overline{R}^\asym = \sqrt{2} G_F [{\bN}^\asym_{ee} - {N}^\asym_{xx}] {R}^\asym \sin[\Delta {S}^\asym] \, , \label{eq:fullasym_Rbar} \\
    &\frac{{R}^\asym}{\overline{R}^\asym} = \begin{cases} \dfrac{{N}^\asym_{ee}-{N}^\asym_{xx}}{{\bN}^\asym_{ee}-{N}^\asym_{xx}} & \text{(minus mode),} \\
    1 & \text{(plus mode).}
    \end{cases} \label{eq:fullasym_eigvec}
\end{align}
\end{subequations}
The ratio of Eqs.~\eqref{eq:fullasym_R} and \eqref{eq:fullasym_Rbar}, combined with Eq.~\eqref{eq:fullasym_eigvec}, leads to the instability threshold conditions \eqref{eq:edge_instab_minus} (minus case) and \eqref{eq:edge_instab_plus} (plus case). Finally, inserting Eq.~\eqref{eq:fullasym_R} in Eq.~\eqref{eq:fullasym_Nee} leads to the prediction for the amount of flavor coherence in the asymptotic state:
\begin{equation}
\label{eq:predict_R}
    R^\asym = \sqrt{- \frac{\kappa_e}{\kappa_e + \kappa_x}[{N}^\asym_{ee}-N_{ee}^{(\mathrm{cl})}][{N}^\asym_{ee}-{N}^\asym_{xx}]} \, .
\end{equation}
The same expression in the antineutrino sector is obtained via Eq.~\eqref{eq:fullasym_eigvec}. We can show that the argument of the square root in Eq.~\eqref{eq:predict_R} is positive, focusing for simplicity on the relevant case $\kappa_x \ll \kappa_e, \, \bkappa_e$. We then have
\begin{equation}
\label{eq:R2}
    \left[R_\pm^\asym\right]^2 \simeq \frac{2 \, a_\pm \, \kappa_x}{(\kappa_e - \bkappa_e)^2} \, \Im(\Omega_\pm) \, \left[N_{ee}^{(\mathrm{cl})} - \bN_{ee}^{(\mathrm{cl})}\right]^2 \, ,
\end{equation}
with $a_+ = 1$ and $a_- = \kappa_e/\bkappa_e$, and the growth rate of the instability is given by Eq.~\eqref{eq:LSA_CFI}. Since by definition, $\Im(\Omega_{\pm}) > 0$ if there is an instability, this ensures that the right-hand side of Eq.~\eqref{eq:R2} is positive.

The prediction~\eqref{eq:predict_R} is shown on the middle panels of Fig.~\ref{fig:pointsAB}, in perfect agreement with the numerical solution of the QKEs. Importantly, these final values are not vanishingly small, which would have been the case with other versions of the collision term (see Sec.~\ref{sec:collision_term}).

\subsection{Duration of relaxation} 
\label{subsec:duration}

We can estimate the time required for the system to relax into the long-term quasi-steady state as follows. In the first phase post-saturation, which occurs on a timescale $\propto \Im(\Omega)^{-1}$ [see Eq.~\eqref{eq:LSA_CFI}], $\nu_x$ adjusts to either $\nu_e$ or $\bar{\nu}_e$, which are themselves brought back approximately to their equilibrium values. We thus write $N_{xx}^{(=)}$ the value of $N_{xx}$ at the end of this first phase, specifically,
\begin{equation}
    N_{xx}^{(=)} \simeq \begin{cases}
    N_{ee}^{(\mathrm{cl})} & \text{minus mode (Point A),} \\
    \bN_{ee}^{(\mathrm{cl})} & \text{plus mode (Point B).}
    \end{cases}
\end{equation}
We crudely describe the next phase of the post-saturation period as an exponential evolution, at the rate $\kappa_x$, of $N_{xx}$ from $N_{xx}^{(=)}$ at $t=t_\mathrm{sat}$ toward the classical steady-state $N_{xx}^{(\mathrm{cl})}$,
\begin{equation}
\label{eq:Nxx_firstphase}
    N_{xx}(t) \simeq \left[N_{xx}^{(=)} - N_{xx}^{(\mathrm{cl})}\right] e^{- \kappa_x(t-t_\mathrm{sat})} + N_{xx}^{(\mathrm{cl})} \, ,
\end{equation}
but this classical relaxation brings the system back into an unstable configuration (see bottom panel of Fig.~\ref{fig:pointsAB}). Flavor coherence grows and flavor conversion $\{\nu_e,\bar{\nu}_e\}\to \nu_x$ occurs. Equation~\eqref{eq:Nxx_firstphase} describes the evolution of $N_{xx}$ only up to a point, which we take to be roughly symmetric to $N_{xx}^{(=)}$ with respect to $N_{xx}^\asym$, that is,\footnote{We use Eq.~\eqref{eq:Nxx_firstphase} from $N_{xx}^{(=)} = N_{xx}^\asym + \Delta N$ to $N_{xx}^\asym - \Delta N$, the latter being therefore equal to $2 N_{xx}^\asym - N_{xx}^{(=)}$.} $2 N_{xx}^{(\infty)} - N_{xx}^{(=)}$. The duration of this phase is thus\footnote{We approximate this portion of exponential relaxation at the rate $\kappa_x$ by its linear counterpart, i.e., $e^{-\kappa_x \Delta t} \simeq 1 - \kappa_x \Delta t$.}
\begin{equation}
\label{eq:timescale}
    \Delta t \sim \frac{2}{\kappa_x} \frac{N_{xx}^{(=)} - N_{xx}^{(\infty)}}{N_{xx}^{(=)} - N_{xx}^{(\mathrm{cl})}} \, ,
\end{equation}
which gives $\Delta t \simeq 180 \, \mathrm{\mu s}$ for both points A and B. This value is consistent with the duration of the first “classical relaxation/instability” cycle in Fig.~\ref{fig:pointsAB}. The overall timescale associated to the evolution toward $N^\asym$ corresponds to a few of these cycles, and can thus be estimated by a few $\Delta t$. One should note, however, that these are long timescales on which the matter field could evolve, such that our effectively homogeneous treatment (see discussion at the end of Sec.~\ref{subsec:model}) could be challenged.

\section{Importance of classical relaxation} 
\label{sec:collision_term}

\begin{figure*}[!ht]
    \centering
    \includegraphics[width=\columnwidth]{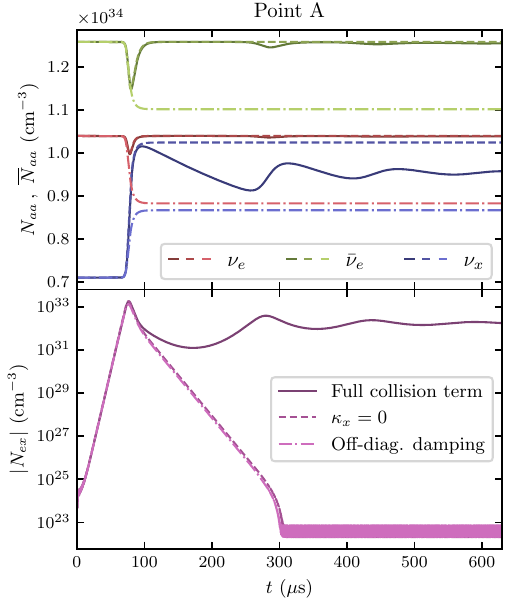}
    \includegraphics[width=\columnwidth]{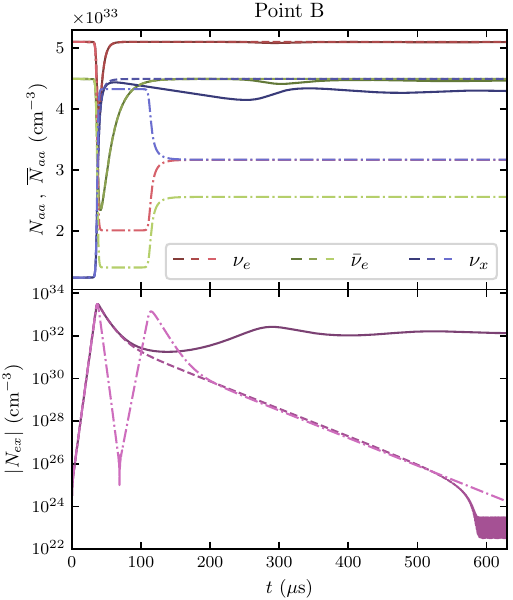}
    \caption{Number density evolution for Point A (left) and B (right), with different treatments of the collision term: full term as in Eq.~\eqref{eq:QKE} (solid lines, same as Fig.~\ref{fig:pointsAB}), $\nu_x$ opacity neglected $\kappa_x = 0$ (dashed lines), and no diagonal repopulation (dash-dotted lines). Flavor coherence is shown on the bottom panels, with only the neutrino off-diagonal component for clarity.}
    \label{fig:compare_collisions_full}
\end{figure*}

A key result of our study is the significant difference of final state when one includes the on-diagonal components of the collision term. To illustrate this, we plot on Fig.~\ref{fig:compare_collisions_full} the evolution of the system for Points A (left) and B (right) under different assumptions: solid lines for the baseline case, dashed lines taking $\kappa_x = \bkappa_x = 0$ (as in, e.g.,~\cite{Johns:2021qby,Fiorillo:2023ajs}), and dash-dotted lines keeping only the flavor off-diagonal collision term (as in, e.g.,~\cite{Zaizen:2025ptx}).

For both points, if the $xx$ entry of the collision term is zero (whether $\kappa_x = 0$ or whether there is only off-diagonal damping), the final amount of flavor coherence is zero up to a small contribution due to the vacuum term. If the on-diagonal terms are kept for $\nu_e$ and $\bar{\nu}_e$, Eq.~\eqref{eq:asym_1} shows that the asymptotic state is $N_{ee} = N_{ee}^{(\mathrm{cl})}$ and $\bN_{ee}=\bN_{ee}^{(\mathrm{cl})}$. For Point A, the final value of $N_{xx}$ is dictated by an approximate flavor “equalization” with $\nu_e$. If $\kappa_x = 0$, then $N_{xx} \simeq N_{ee}^{(\mathrm{cl})} = N_{xx}^{(=)}$, consistently with the discussion in Sec.~\ref{subsec:duration} (since the “true” asymptotic state $N_{xx}^{(\infty)}$ is pushed to $t \to + \infty$ as $\kappa_x \to 0$). If there is only flavor off-diagonal damping, we recover the usual flavor equipartition discussed in other works (see, e.g.,~\cite{Lin:2022dek,Johns:2023xae,Zaizen:2025ptx}), and $\bN_{ee}$ is set by the conservation of $\Tr(\bN)$ (since there are no on-diagonal collision terms that can change the trace).

For Point B, the evolution in the $\kappa_x = 0$ case is similar to that of Point A: after the saturation of the instability, $N_{xx}$ remains stuck on $N_{xx}^{(=)} = \bN_{ee}^{(\mathrm{cl})}$, and $N_{ee}$ and $\bN_{ee}$ are back to their classical equilibrium values. The evolution in the “off-diagonal damping only” case is more subtle. At first, following the plus instability, there is a flavor swap between $\nu_x$ and $\bar{\nu}_e$ (see the almost perfect inversion of the dash-dotted blue and green lines) in the first $100 \, \mathrm{\mu s}$. However, this new configuration is unstable, but for the \emph{minus} mode, while the plus mode is stable (with a negative imaginary part). This is the reason why $|N_{ex}|$ first decreases, as the plus mode decays, before the newly unstable minus mode takes over and leads to a second instability, this time inducing the equipartition of $\nu_x$ and $\nu_e$. With the “full” collision term, this change of unstable mode is prevented by the fact that $N_{ee}$ and $\bN_{ee}$ remain close to their classical equilibrium values. Interestingly, the outcome (swap/equipartition) of the “off-diagonal damping only” CFI is related to which mode (plus/minus) is unstable, as noted in~\cite{Zaizen:2025ptx}, although that work used multienergy distributions.

Importantly, there is no flavor equipartition when the repopulation of $\nu_e$ and $\bar{\nu}_e$ is taken into account, as their densities settle close (or exactly, if $\kappa_x = 0$) to their classical equilibrium values, far from the result obtained without diagonal collision terms, in dash-dotted lines. This incomplete treatment misses, in the cases we look at, the net pair production of $\nu_x$ through the combination of $\nu_e/\bar{\nu}_e$ repopulation and oscillations.

\section{Outlook} 
\label{sec:outlook}

We have studied the non-linear evolution of a neutrino/antineutrino system encountering a collisional flavor instability, deriving analytical expressions for the quasi-steady state. Notably, we have shown that studying CFIs in a setup where the effect of collisions is simply to damp the off-diagonal flavor coherence leads to incorrect asymptotic values, as it overlooks the main effect of collisions which are to drive number densities back toward their classical equilibrium values. On short timescales ($\propto \kappa_{e}^{-1}, \bkappa_{e}^{-1}$), $\nu_x$ adjusts to match $\nu_e$ or $\bar{\nu}_e$, which are close to their classical equilibriums, thus resulting in a net emission of heavy-lepton flavor (anti)neutrinos. Including the classical relaxation toward equilibrium of $\nu_e$ and $\bar{\nu}_e$ prevents the flavor equipartition observed in other works, see the difference of solid and dashed lines with dash-dotted lines on Fig.~\ref{fig:compare_collisions_full}. Following the saturation of the first instability, competition between instability of and relaxation toward the classical steady-state leads to a “compromise” where the system lies at the edge of collisional instability, thus defining a “quantum” equilibrium. This asymptotic state is described analytically by Eqs.~\eqref{eq:asym_1}--\eqref{eq:Neeasym_plus} and \eqref{eq:predict_R}, with distinctions depending on the branch (“plus” or “minus”) of the instability. We highlight that flavor coherence is nonzero in this final state, contrary to what could naively be expected from collisions. It is precisely this nonzero flavor coherence that allows the system to sustain the difference between classical and quantum steady-states $N_{aa}^{(\mathrm{cl})} \neq N_{aa}^{(\infty)}$. This work provides the first explicit prediction of this asymptotic state that could be used in a subgrid model of CFIs, in spite of limitations that will need to be addressed in the future. 

First, we have considered monochromatic systems, thus missing the phenomenology associated with the energy dependence of collision rates (see e.g.,~\cite{Zaizen:2025ptx}). We have also restricted our study to homogeneous and isotropic systems. Although isotropy is well-justified in the regions where CFIs will matter, inhomogeneous effects could be important (see e.g.,~\cite{Bhattacharyya:2025gds} for a recent study in the fast regime), especially given the large timescale of relaxation evidenced here. Notably, we have considered an effective homogeneous relaxation toward the classical steady-state (especially important for the heavy-lepton flavor neutrinos), which should be compared with larger-scale inhomogeneous simulations, which we leave for future work. These limitations form as many future directions for a more general description of CFIs in dense environments.

\begin{acknowledgments}
    
I thank J. Kneller, G. McLaughlin, S. Richers, F. Foucart, E. Grohs, C. Hall and E. Urquilla for many useful discussions, and in particular J. Kneller, G. McLaughlin, S. Richers, H. Nagakura, M. Zaizen and D. Fiorillo for their valuable comments on the manuscript. This work was supported by the Network for Neutrinos, Nuclear Astrophysics and Symmetries (N3AS), through the National Science Foundation Physics Frontier Center award No. PHY-2020275.

\end{acknowledgments}

\bigskip

\appendix

\section{Linear stability analysis}
\label{app:LSA}

We detail here the results of the linear stability analysis of the system described by Eq.~\eqref{eq:QKE}. 

To make the connection with other works in the literature, specifically~\cite{Liu:2023pjw,Liu:2023vtz,Akaho:2023brj,Liu:2024wzd,Zaizen:2024faj,Zaizen:2025ptx}, we introduce the notations
\begin{equation}
\label{eq:notations_Zaizen}
\begin{aligned}
    G &\equiv \frac{\mathfrak{g} + \bar{\mathfrak{g}}}{2} \, ,  & \qquad \gamma &\equiv \frac{\Gamma + \overline{\Gamma}}{2} \, , \\
    A &\equiv \frac{\mathfrak{g} - \bar{\mathfrak{g}}}{2} \, ,  & \qquad \alpha &\equiv \frac{\Gamma - \overline{\Gamma}}{2} \, .
\end{aligned}
\end{equation}
%
%

\begin{widetext}

The stability matrix in Eq.~\eqref{eq:QKE_linear} can be rewritten
\begin{equation}
    \Lambda = \begin{bmatrix} A - G - \i \, (\gamma + \alpha) - \omega \, \ct + \lambda & A +G \\
    A - G & A + G - \i \, (\gamma - \alpha) + \omega \, \ct + \lambda \end{bmatrix} \, .
\end{equation}
Its eigenvalues and eigenvectors are\footnote{Some references define shifted frequencies $\Omega' \equiv \Omega - 2A - \lambda$, such that the real parts of the eigenvalues can appear different between various sources, while the imaginary parts remain the same.}
\begin{equation}
\begin{aligned}
    Q_1 &= \begin{bmatrix} G + \i \, \alpha_\omega + \sqrt{A^2 + 2 \, \i \, G \, \alpha_\omega - \alpha_\omega^2} \\ G-A \end{bmatrix} \, , & \  \Omega_1 &= \lambda + A - \sqrt{A^2 + 2 \, \i \, G \, \alpha_\omega - \alpha_\omega^2} - \i \, \gamma \, ,   \\
    Q_2 &= \begin{bmatrix} G + \i \, \alpha_\omega - \sqrt{A^2 + 2 \, \i \, G \, \alpha_\omega - \alpha_\omega^2} \\ G-A \end{bmatrix} \, ,
      & \ \Omega_2 &= \lambda + A + \sqrt{A^2 + 2 \, \i \, G \, \alpha_\omega - \alpha_\omega^2} - \i \, \gamma \, ,
\end{aligned}
\end{equation}
where we wrote $\i \alpha_\omega \equiv \i \alpha + \omega \ct$. Outside of the resonancelike regime, we have $\lvert G \alpha \rvert \ll A^2$ (and also $|\alpha| \ll G$, $|\omega \ct| \ll G$), which allows one to simplify the expressions as

\begin{equation}
\label{eq:LSA_12}
\begin{aligned}
    Q_1 &\simeq \begin{bmatrix} G + |A| 
    \\ G - A \end{bmatrix} \, , & \qquad \quad  \Omega_1 &\simeq \lambda + (A - |A|) - \frac{G \omega \ct}{|A|} + \i \left(-\gamma - \frac{G \alpha}{|A|}\right) \, ,  \\
     \qquad Q_2 &\simeq \begin{bmatrix} G - |A| 
    \\ G - A \end{bmatrix} \, , & \qquad \quad \Omega_2 &\simeq \lambda + (A + |A|) + \frac{G \omega \ct}{|A|} + \i \left(- \gamma + \frac{G \alpha}{|A|}\right) \, .
\end{aligned}
\end{equation}
\end{widetext}

Depending on the sign of the various quantities, we always have an eigenmode (the “minus” mode)
\begin{equation}
\label{eq:Q_minus}
    Q_- \simeq \begin{bmatrix} G + A \\ G - A \end{bmatrix} = \begin{bmatrix} \mathfrak{g} \\ \bar{\mathfrak{g}} \end{bmatrix} \, ,
\end{equation}
with eigenvalue
\begin{equation}
\label{eq:Om_minus}
\begin{aligned}
    \Omega_- &\simeq \lambda - \frac{G \omega \ct}{A} + \i \left(- \gamma - \frac{G \alpha}{A} \right) \\
    &= \lambda - \frac{\mathfrak{g} + \bar{\mathfrak{g}}}{\mathfrak{g} - \bar{\mathfrak{g}}}\omega \ct + \i \frac{\overline{\Gamma} \bar{\mathfrak{g}} - \Gamma \mathfrak{g}}{\mathfrak{g} - \bar{\mathfrak{g}}} \, ,
\end{aligned}
\end{equation}
in agreement with Eq.~(14) in~\cite{Lin:2022dek}. The other eigenmode (the “plus” mode) is, after multiplying the eigenvector by $(G-A)^{-1}$,
\begin{equation}
\label{eq:Q_plus}
Q_+ \simeq \begin{bmatrix} 1 \\ 1 \end{bmatrix} \, ,
\end{equation}
with eigenvalue
\begin{equation}
\label{eq:Om_plus}
\begin{aligned}
    \Omega_+ &\simeq \lambda + \frac{G \omega \ct}{A} + 2 A + \i \left(- \gamma + \frac{G \alpha}{A} \right) \\
    &= \lambda + \frac{\mathfrak{g} + \bar{\mathfrak{g}}}{\mathfrak{g} - \bar{\mathfrak{g}}}\omega \ct + \mathfrak{g} - \bar{\mathfrak{g}} + \i \frac{\Gamma \bar{\mathfrak{g}} - \overline{\Gamma} \mathfrak{g}}{\mathfrak{g} - \bar{\mathfrak{g}}} \, ,
\end{aligned}
\end{equation}
in agreement with Eq.~(13) in~\cite{Lin:2022dek}. The imaginary parts of Eqs.~\eqref{eq:Om_minus} and~\eqref{eq:Om_plus} are given in the main text, Eq.~\eqref{eq:LSA_CFI}.

We decide here to designate the “plus/minus” mode depending on the expression of the eigenvector. Indeed, this is a \emph{physical} distinction (see e.g.,~\cite{Fiorillo:2023ajs,Johns:2025mlm} for further discussions). In the polarization vector formalism,\footnote{We can represent the $2\times 2$ matrix $N = \frac12 \left(P_0 \, \mathbb{I} + \mathbf{P} \cdot \bm{\sigma}\right)$, with $\bm{\sigma} = \left(\sigma_x, \sigma_y, \sigma_z\right)^T$ the “vector” of Pauli matrices. $\mathbf{P}$ is the “polarization vector.”} the minus mode corresponds to a collective transverse growth of $\mathbf{P}$ and $\mathbf{\bP}$, which are “locked” on $\mathbf{P} - \mathbf{\bP}$; in other words their common axis of precession tilts away from $\hat{\mathbf{z}}$. For the plus mode, the axis of precession remains essentially $\hat{\mathbf{z}}$, but the aperture of the precession cones grows exponentially, with $\mathbf{P}$ and $\overline{\mathbf{P}}$ “blooming” and developing large transverse components.

\begin{figure}[!ht]
    \centering
    \includegraphics[width=\columnwidth]{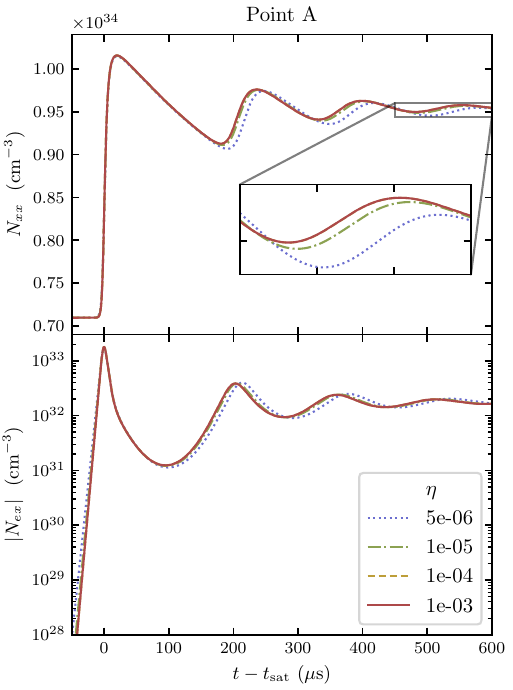}
    \caption{Modification of the evolution for different attenuation factors $\eta$ in the Hamiltonian (see text). There is no discernible difference for attenuations larger than $10^{-4}$, where we are safely in the regime $\Gamma \ll ||H||$. For comparison purposes, the calculations are aligned at the saturation time $t_\mathrm{sat}$, which corresponds to the first peak in $|N_{ex}|$.}
    \label{fig:attenuation_factor}
\end{figure}

\section{Attenuation of the Hamiltonian} 
\label{app:attenuation}

As mentioned in Sec.~\ref{subsec:numerical}, for computational reasons we use an attenuated Hamiltonian, where the vacuum, matter and self-interaction contributions in Eq.~\eqref{eq:QKE} are multiplied by an attenuation factor\footnote{This approach is slightly different from the one in, e.g.,~\cite{Xiong:2022vsy,Shalgar:2024gjt}, where only the self-interaction Hamiltonian $H_{\nu \nu}$ is rescaled.} $\eta < 1$. This is necessary in practice because the matter term in particular introduces an extremely small timescale (see Table~\ref{tab:timescales}). 
In addition, the matter term reduces the amplitude of the eigenvectors $Q_\pm$ in Eq.~\eqref{eq:Nex_sol}, see Eq.~(12) in \cite{Lin:2022dek} with the expressions of $\Omega_\pm$ given in Eqs.~\eqref{eq:Om_minus} and~\eqref{eq:Om_plus}, thus requiring exceedingly small error tolerances in the differential equation solver. 

\renewcommand{\arraystretch}{1.3}

\begin{table}[!ht]
    \centering
    \setlength{\tabcolsep}{8pt}
    \begin{tabular}{|r|c|c|}
    \hline
        & Point A & Point B \\ \hline \hline
        $\sqrt{2} G_F |N_{ee} - \bN_{ee}| \ (\mathrm{s}^{-1})$ & $4.2 \times 10^{11}$ & $1.2 \times 10^{11}$ \\
        $\sqrt{2} G_F n_e \ (\mathrm{s}^{-1})$ & $1.0 \times 10^{13}$  & $6.9 \times 10^{12}$ \\
        $\Delta m^2 / 2 \langle E_\nu \rangle \ (\mathrm{s}^{-1})$ & $5.3 \times 10^4$ & $5.5 \times 10^4$ \\ \hline
    \end{tabular}
    \caption{Frequencies involved in Eq.~\eqref{eq:QKE} for the conditions of the points in Table~\ref{tab:NSMpoints}, where the classical relaxation timescales are reported.
    }
    \label{tab:timescales}
\end{table}

We observe from Fig.~\ref{fig:attenuation_factor} that using $\eta > 10^{-4}$ does not change the evolution and in particular the final state equilibrium; we are then safely in the regime where the classical relaxation timescale is much larger than the self-interaction (and matter) timescales. Using the notations from Appendix~\ref{app:LSA}, attenuating the Hamiltonian by $\eta$ corresponds to rescaling $G \to \eta G$ and $A \to \eta A$. Therefore, as long as $\lvert (\eta G) \alpha\rvert \ll (\eta A)^2$, the instability growth rate is given by the approximate Eq.~\eqref{eq:LSA_12} where the factors of $\eta$ cancel out, which explains why the growth rate is unchanged in Fig.~\ref{fig:attenuation_factor}. We have checked that smaller values $\eta < 10^{-6}$ (which break the physical ordering of timescales) lead to a different evolution, with in particular a smaller growth rate.

Ideally, one could circumvent this attenuation procedure by tracking only the “slow” evolution driven by collisions (which have secular effects on the very fast oscillations happening on matter/self-interaction timescales), following strategies which leverage the large timescale separation, such as the adiabatic-like approximation in~\cite{Froustey:2020mcq,Froustey:2021azz}, the coarse graining of~\cite{Johns:2023xae} or the “slow-dynamics equations” derived in~\cite{Fiorillo:2023ajs}.

\bibliography{references}

\begin{thebibliography}{88}%
\makeatletter
\providecommand \@ifxundefined [1]{%
 \@ifx{#1\undefined}
}%
\providecommand \@ifnum [1]{%
 \ifnum #1\expandafter \@firstoftwo
 \else \expandafter \@secondoftwo
 \fi
}%
\providecommand \@ifx [1]{%
 \ifx #1\expandafter \@firstoftwo
 \else \expandafter \@secondoftwo
 \fi
}%
\providecommand \natexlab [1]{#1}%
\providecommand \enquote  [1]{``#1''}%
\providecommand \bibnamefont  [1]{#1}%
\providecommand \bibfnamefont [1]{#1}%
\providecommand \citenamefont [1]{#1}%
\providecommand \href@noop [0]{\@secondoftwo}%
\providecommand \href [0]{\begingroup \@sanitize@url \@href}%
\providecommand \@href[1]{\@@startlink{#1}\@@href}%
\providecommand \@@href[1]{\endgroup#1\@@endlink}%
\providecommand \@sanitize@url [0]{\catcode `\\12\catcode `\$12\catcode `\&12\catcode `\#12\catcode `\^12\catcode `\_12\catcode `\%12\relax}%
\providecommand \@@startlink[1]{}%
\providecommand \@@endlink[0]{}%
\providecommand \url  [0]{\begingroup\@sanitize@url \@url }%
\providecommand \@url [1]{\endgroup\@href {#1}{\urlprefix }}%
\providecommand \urlprefix  [0]{URL }%
\providecommand \Eprint [0]{\href }%
\providecommand \doibase [0]{https://doi.org/}%
\providecommand \selectlanguage [0]{\@gobble}%
\providecommand \bibinfo  [0]{\@secondoftwo}%
\providecommand \bibfield  [0]{\@secondoftwo}%
\providecommand \translation [1]{[#1]}%
\providecommand \BibitemOpen [0]{}%
\providecommand \bibitemStop [0]{}%
\providecommand \bibitemNoStop [0]{.\EOS\space}%
\providecommand \EOS [0]{\spacefactor3000\relax}%
\providecommand \BibitemShut  [1]{\csname bibitem#1\endcsname}%
\let\auto@bib@innerbib\@empty
\bibitem [{\citenamefont {Janka}\ \emph {et~al.}(2016)\citenamefont {Janka}, \citenamefont {Melson},\ and\ \citenamefont {Summa}}]{Janka:2016fox}%
  \BibitemOpen
  \bibfield  {author} {\bibinfo {author} {\bibfnamefont {H.~T.}\ \bibnamefont {Janka}}, \bibinfo {author} {\bibfnamefont {T.}~\bibnamefont {Melson}},\ and\ \bibinfo {author} {\bibfnamefont {A.}~\bibnamefont {Summa}},\ }\bibfield  {title} {\bibinfo {title} {{Physics of Core-Collapse Supernovae in Three Dimensions: a Sneak Preview}},\ }\href {https://doi.org/10.1146/annurev-nucl-102115-044747} {\bibfield  {journal} {\bibinfo  {journal} {Ann. Rev. Nucl. Part. Sci.}\ }\textbf {\bibinfo {volume} {66}},\ \bibinfo {pages} {341} (\bibinfo {year} {2016})},\ \Eprint {https://arxiv.org/abs/1602.05576} {arXiv:1602.05576 [astro-ph.SR]} \BibitemShut {NoStop}%
\bibitem [{\citenamefont {Burrows}\ and\ \citenamefont {Vartanyan}(2021)}]{Burrows:2020qrp}%
  \BibitemOpen
  \bibfield  {author} {\bibinfo {author} {\bibfnamefont {A.}~\bibnamefont {Burrows}}\ and\ \bibinfo {author} {\bibfnamefont {D.}~\bibnamefont {Vartanyan}},\ }\bibfield  {title} {\bibinfo {title} {{Core-Collapse Supernova Explosion Theory}},\ }\href {https://doi.org/10.1038/s41586-020-03059-w} {\bibfield  {journal} {\bibinfo  {journal} {Nature}\ }\textbf {\bibinfo {volume} {589}},\ \bibinfo {pages} {29} (\bibinfo {year} {2021})},\ \Eprint {https://arxiv.org/abs/2009.14157} {arXiv:2009.14157 [astro-ph.SR]} \BibitemShut {NoStop}%
\bibitem [{\citenamefont {Kyutoku}\ \emph {et~al.}(2021)\citenamefont {Kyutoku}, \citenamefont {Shibata},\ and\ \citenamefont {Taniguchi}}]{Kyutoku:2021icp}%
  \BibitemOpen
  \bibfield  {author} {\bibinfo {author} {\bibfnamefont {K.}~\bibnamefont {Kyutoku}}, \bibinfo {author} {\bibfnamefont {M.}~\bibnamefont {Shibata}},\ and\ \bibinfo {author} {\bibfnamefont {K.}~\bibnamefont {Taniguchi}},\ }\bibfield  {title} {\bibinfo {title} {{Coalescence of black hole-neutron star binaries}},\ }\href {https://doi.org/10.1007/s41114-021-00033-4} {\bibfield  {journal} {\bibinfo  {journal} {Living Rev. Rel.}\ }\textbf {\bibinfo {volume} {24}},\ \bibinfo {pages} {5} (\bibinfo {year} {2021})},\ \Eprint {https://arxiv.org/abs/2110.06218} {arXiv:2110.06218 [astro-ph.HE]} \BibitemShut {NoStop}%
\bibitem [{\citenamefont {Radice}\ \emph {et~al.}(2020)\citenamefont {Radice}, \citenamefont {Bernuzzi},\ and\ \citenamefont {Perego}}]{Radice:2020ddv}%
  \BibitemOpen
  \bibfield  {author} {\bibinfo {author} {\bibfnamefont {D.}~\bibnamefont {Radice}}, \bibinfo {author} {\bibfnamefont {S.}~\bibnamefont {Bernuzzi}},\ and\ \bibinfo {author} {\bibfnamefont {A.}~\bibnamefont {Perego}},\ }\bibfield  {title} {\bibinfo {title} {{The Dynamics of Binary Neutron Star Mergers and GW170817}},\ }\href {https://doi.org/10.1146/annurev-nucl-013120-114541} {\bibfield  {journal} {\bibinfo  {journal} {Ann. Rev. Nucl. Part. Sci.}\ }\textbf {\bibinfo {volume} {70}},\ \bibinfo {pages} {95} (\bibinfo {year} {2020})},\ \Eprint {https://arxiv.org/abs/2002.03863} {arXiv:2002.03863 [astro-ph.HE]} \BibitemShut {NoStop}%
\bibitem [{\citenamefont {Mezzacappa}(2020)}]{Mezzacappa:2020pkk}%
  \BibitemOpen
  \bibfield  {author} {\bibinfo {author} {\bibfnamefont {A.}~\bibnamefont {Mezzacappa}},\ }\bibfield  {title} {\bibinfo {title} {{Toward Realistic Models of Core Collapse Supernovae: A Brief Review}},\ }\href {https://doi.org/10.1017/S1743921322001831} {\bibfield  {journal} {\bibinfo  {journal} {IAU Symp.}\ }\textbf {\bibinfo {volume} {362}},\ \bibinfo {pages} {215} (\bibinfo {year} {2020})},\ \Eprint {https://arxiv.org/abs/2205.13438} {arXiv:2205.13438 [astro-ph.SR]} \BibitemShut {NoStop}%
\bibitem [{\citenamefont {Kiuchi}(2025)}]{Kiuchi:2024lpx}%
  \BibitemOpen
  \bibfield  {author} {\bibinfo {author} {\bibfnamefont {K.}~\bibnamefont {Kiuchi}},\ }\bibinfo {title} {General relativistic magnetohydrodynamics simulations for binary neutron star mergers},\ in\ \href {https://doi.org/10.1007/978-981-97-8522-3_16} {\emph {\bibinfo {booktitle} {New Frontiers in GRMHD Simulations}}},\ \bibinfo {editor} {edited by\ \bibinfo {editor} {\bibfnamefont {C.}~\bibnamefont {Bambi}}, \bibinfo {editor} {\bibfnamefont {Y.}~\bibnamefont {Mizuno}}, \bibinfo {editor} {\bibfnamefont {S.}~\bibnamefont {Shashank}},\ and\ \bibinfo {editor} {\bibfnamefont {F.}~\bibnamefont {Yuan}}}\ (\bibinfo  {publisher} {Springer Nature Singapore},\ \bibinfo {address} {Singapore},\ \bibinfo {year} {2025})\ pp.\ \bibinfo {pages} {529--572},\ \Eprint {https://arxiv.org/abs/2405.10081} {arXiv:2405.10081 [astro-ph.HE]} \BibitemShut {NoStop}%
\bibitem [{\citenamefont {Mezzacappa}\ \emph {et~al.}(2020)\citenamefont {Mezzacappa}, \citenamefont {Endeve}, \citenamefont {Bronson~Messer},\ and\ \citenamefont {Bruenn}}]{Mezzacappa:2020oyq}%
  \BibitemOpen
  \bibfield  {author} {\bibinfo {author} {\bibfnamefont {A.}~\bibnamefont {Mezzacappa}}, \bibinfo {author} {\bibfnamefont {E.}~\bibnamefont {Endeve}}, \bibinfo {author} {\bibfnamefont {O.~E.}\ \bibnamefont {Bronson~Messer}},\ and\ \bibinfo {author} {\bibfnamefont {S.~W.}\ \bibnamefont {Bruenn}},\ }\bibfield  {title} {\bibinfo {title} {{Physical, numerical, and computational challenges of modeling neutrino transport in core-collapse supernovae}},\ }\href {https://doi.org/10.1007/s41115-020-00010-8} {\bibfield  {journal} {\bibinfo  {journal} {Living Rev. Comput. Astrophys.}\ }\textbf {\bibinfo {volume} {6}},\ \bibinfo {pages} {4} (\bibinfo {year} {2020})},\ \Eprint {https://arxiv.org/abs/2010.09013} {arXiv:2010.09013 [astro-ph.HE]} \BibitemShut {NoStop}%
\bibitem [{\citenamefont {Foucart}(2023)}]{Foucart:2022bth}%
  \BibitemOpen
  \bibfield  {author} {\bibinfo {author} {\bibfnamefont {F.}~\bibnamefont {Foucart}},\ }\bibfield  {title} {\bibinfo {title} {{Neutrino transport in general relativistic neutron star merger simulations}},\ }\href {https://doi.org/10.1007/s41115-023-00016-y} {\bibfield  {journal} {\bibinfo  {journal} {Living Rev. Comput. Astrophys.}\ }\textbf {\bibinfo {volume} {9}},\ \bibinfo {pages} {1} (\bibinfo {year} {2023})},\ \Eprint {https://arxiv.org/abs/2209.02538} {arXiv:2209.02538 [astro-ph.HE]} \BibitemShut {NoStop}%
\bibitem [{\citenamefont {Fischer}\ \emph {et~al.}(2024)\citenamefont {Fischer}, \citenamefont {Guo}, \citenamefont {Langanke}, \citenamefont {Martinez-Pinedo}, \citenamefont {Qian},\ and\ \citenamefont {Wu}}]{Fischer:2023ebq}%
  \BibitemOpen
  \bibfield  {author} {\bibinfo {author} {\bibfnamefont {T.}~\bibnamefont {Fischer}}, \bibinfo {author} {\bibfnamefont {G.}~\bibnamefont {Guo}}, \bibinfo {author} {\bibfnamefont {K.}~\bibnamefont {Langanke}}, \bibinfo {author} {\bibfnamefont {G.}~\bibnamefont {Martinez-Pinedo}}, \bibinfo {author} {\bibfnamefont {Y.-Z.}\ \bibnamefont {Qian}},\ and\ \bibinfo {author} {\bibfnamefont {M.-R.}\ \bibnamefont {Wu}},\ }\bibfield  {title} {\bibinfo {title} {{Neutrinos and nucleosynthesis of elements}},\ }\href {https://doi.org/10.1016/j.ppnp.2024.104107} {\bibfield  {journal} {\bibinfo  {journal} {Prog. Part. Nucl. Phys.}\ }\textbf {\bibinfo {volume} {137}},\ \bibinfo {pages} {104107} (\bibinfo {year} {2024})},\ \Eprint {https://arxiv.org/abs/2308.03962} {arXiv:2308.03962 [astro-ph.HE]} \BibitemShut {NoStop}%
\bibitem [{\citenamefont {Wang}\ and\ \citenamefont {Surman}(2023)}]{Wang:2023tso}%
  \BibitemOpen
  \bibfield  {author} {\bibinfo {author} {\bibfnamefont {X.}~\bibnamefont {Wang}}\ and\ \bibinfo {author} {\bibfnamefont {R.}~\bibnamefont {Surman}},\ }\bibinfo {title} {{Neutrinos and Heavy Element Nucleosynthesis}},\ in\ \href {https://doi.org/10.1007/978-981-15-8818-1_128-1} {\emph {\bibinfo {booktitle} {{Handbook of Nuclear Physics}}}},\ \bibinfo {editor} {edited by\ \bibinfo {editor} {\bibfnamefont {I.}~\bibnamefont {Tanihata}}, \bibinfo {editor} {\bibfnamefont {H.}~\bibnamefont {Toki}},\ and\ \bibinfo {editor} {\bibfnamefont {T.}~\bibnamefont {Kajino}}}\ (\bibinfo  {publisher} {Springer Nature Singapore},\ \bibinfo {address} {Singapore},\ \bibinfo {year} {2023})\ pp.\ \bibinfo {pages} {1--19},\ \Eprint {https://arxiv.org/abs/2309.06043} {arXiv:2309.06043 [astro-ph.HE]} \BibitemShut {NoStop}%
\bibitem [{\citenamefont {Duan}\ \emph {et~al.}(2010)\citenamefont {Duan}, \citenamefont {Fuller},\ and\ \citenamefont {Qian}}]{Duan:2010bg}%
  \BibitemOpen
  \bibfield  {author} {\bibinfo {author} {\bibfnamefont {H.}~\bibnamefont {Duan}}, \bibinfo {author} {\bibfnamefont {G.~M.}\ \bibnamefont {Fuller}},\ and\ \bibinfo {author} {\bibfnamefont {Y.-Z.}\ \bibnamefont {Qian}},\ }\bibfield  {title} {\bibinfo {title} {{Collective Neutrino Oscillations}},\ }\href {https://doi.org/10.1146/annurev.nucl.012809.104524} {\bibfield  {journal} {\bibinfo  {journal} {Annu. Rev. Nucl. Part. Sci.}\ }\textbf {\bibinfo {volume} {60}},\ \bibinfo {pages} {569} (\bibinfo {year} {2010})},\ \Eprint {https://arxiv.org/abs/1001.2799} {arXiv:1001.2799 [hep-ph]} \BibitemShut {NoStop}%
\bibitem [{\citenamefont {Mirizzi}\ \emph {et~al.}(2016)\citenamefont {Mirizzi}, \citenamefont {Tamborra}, \citenamefont {Janka}, \citenamefont {Saviano}, \citenamefont {Scholberg}, \citenamefont {Bollig}, \citenamefont {Hudepohl},\ and\ \citenamefont {Chakraborty}}]{Mirizzi:2015eza}%
  \BibitemOpen
  \bibfield  {author} {\bibinfo {author} {\bibfnamefont {A.}~\bibnamefont {Mirizzi}}, \bibinfo {author} {\bibfnamefont {I.}~\bibnamefont {Tamborra}}, \bibinfo {author} {\bibfnamefont {H.-T.}\ \bibnamefont {Janka}}, \bibinfo {author} {\bibfnamefont {N.}~\bibnamefont {Saviano}}, \bibinfo {author} {\bibfnamefont {K.}~\bibnamefont {Scholberg}}, \bibinfo {author} {\bibfnamefont {R.}~\bibnamefont {Bollig}}, \bibinfo {author} {\bibfnamefont {L.}~\bibnamefont {Hudepohl}},\ and\ \bibinfo {author} {\bibfnamefont {S.}~\bibnamefont {Chakraborty}},\ }\bibfield  {title} {\bibinfo {title} {{Supernova Neutrinos: Production, Oscillations and Detection}},\ }\href {https://doi.org/10.1393/ncr/i2016-10120-8} {\bibfield  {journal} {\bibinfo  {journal} {Riv. Nuovo Cim.}\ }\textbf {\bibinfo {volume} {39}},\ \bibinfo {pages} {1} (\bibinfo {year} {2016})},\ \Eprint {https://arxiv.org/abs/1508.00785} {arXiv:1508.00785 [astro-ph.HE]} \BibitemShut {NoStop}%
\bibitem [{\citenamefont {Chakraborty}\ \emph {et~al.}(2016)\citenamefont {Chakraborty}, \citenamefont {Hansen}, \citenamefont {Izaguirre},\ and\ \citenamefont {Raffelt}}]{Chakraborty:2016yeg}%
  \BibitemOpen
  \bibfield  {author} {\bibinfo {author} {\bibfnamefont {S.}~\bibnamefont {Chakraborty}}, \bibinfo {author} {\bibfnamefont {R.}~\bibnamefont {Hansen}}, \bibinfo {author} {\bibfnamefont {I.}~\bibnamefont {Izaguirre}},\ and\ \bibinfo {author} {\bibfnamefont {G.}~\bibnamefont {Raffelt}},\ }\bibfield  {title} {\bibinfo {title} {{Collective neutrino flavor conversion: Recent developments}},\ }\href {https://doi.org/10.1016/j.nuclphysb.2016.02.012} {\bibfield  {journal} {\bibinfo  {journal} {Nucl. Phys. B}\ }\textbf {\bibinfo {volume} {908}},\ \bibinfo {pages} {366} (\bibinfo {year} {2016})},\ \Eprint {https://arxiv.org/abs/1602.02766} {arXiv:1602.02766 [hep-ph]} \BibitemShut {NoStop}%
\bibitem [{\citenamefont {Tamborra}\ and\ \citenamefont {Shalgar}(2021)}]{Tamborra:2020cul}%
  \BibitemOpen
  \bibfield  {author} {\bibinfo {author} {\bibfnamefont {I.}~\bibnamefont {Tamborra}}\ and\ \bibinfo {author} {\bibfnamefont {S.}~\bibnamefont {Shalgar}},\ }\bibfield  {title} {\bibinfo {title} {{New Developments in Flavor Evolution of a Dense Neutrino Gas}},\ }\href {https://doi.org/10.1146/annurev-nucl-102920-050505} {\bibfield  {journal} {\bibinfo  {journal} {Annu. Rev. Nucl. Part. Sci.}\ }\textbf {\bibinfo {volume} {71}},\ \bibinfo {pages} {165} (\bibinfo {year} {2021})},\ \Eprint {https://arxiv.org/abs/2011.01948} {arXiv:2011.01948 [astro-ph.HE]} \BibitemShut {NoStop}%
\bibitem [{\citenamefont {Capozzi}\ and\ \citenamefont {Saviano}(2022)}]{Capozzi:2022slf}%
  \BibitemOpen
  \bibfield  {author} {\bibinfo {author} {\bibfnamefont {F.}~\bibnamefont {Capozzi}}\ and\ \bibinfo {author} {\bibfnamefont {N.}~\bibnamefont {Saviano}},\ }\bibfield  {title} {\bibinfo {title} {{Neutrino Flavor Conversions in High-Density Astrophysical and Cosmological Environments}},\ }\href {https://doi.org/10.3390/universe8020094} {\bibfield  {journal} {\bibinfo  {journal} {Universe}\ }\textbf {\bibinfo {volume} {8}},\ \bibinfo {pages} {94} (\bibinfo {year} {2022})},\ \Eprint {https://arxiv.org/abs/2202.02494} {arXiv:2202.02494 [hep-ph]} \BibitemShut {NoStop}%
\bibitem [{\citenamefont {Volpe}(2024)}]{Volpe:2023met}%
  \BibitemOpen
  \bibfield  {author} {\bibinfo {author} {\bibfnamefont {M.~C.}\ \bibnamefont {Volpe}},\ }\bibfield  {title} {\bibinfo {title} {{Neutrinos from dense environments: Flavor mechanisms, theoretical approaches, observations, and new directions}},\ }\href {https://doi.org/10.1103/RevModPhys.96.025004} {\bibfield  {journal} {\bibinfo  {journal} {Rev. Mod. Phys.}\ }\textbf {\bibinfo {volume} {96}},\ \bibinfo {pages} {025004} (\bibinfo {year} {2024})},\ \Eprint {https://arxiv.org/abs/2301.11814} {arXiv:2301.11814 [hep-ph]} \BibitemShut {NoStop}%
\bibitem [{\citenamefont {Richers}\ and\ \citenamefont {Sen}(2022)}]{Richers_review}%
  \BibitemOpen
  \bibfield  {author} {\bibinfo {author} {\bibfnamefont {S.}~\bibnamefont {Richers}}\ and\ \bibinfo {author} {\bibfnamefont {M.}~\bibnamefont {Sen}},\ }\bibinfo {title} {{Fast Flavor Transformations}},\ in\ \href {https://doi.org/10.1007/978-981-15-8818-1_125-1} {\emph {\bibinfo {booktitle} {{Handbook of Nuclear Physics}}}},\ \bibinfo {editor} {edited by\ \bibinfo {editor} {\bibfnamefont {I.}~\bibnamefont {Tanihata}}, \bibinfo {editor} {\bibfnamefont {H.}~\bibnamefont {Toki}},\ and\ \bibinfo {editor} {\bibfnamefont {T.}~\bibnamefont {Kajino}}}\ (\bibinfo  {publisher} {Springer Nature Singapore},\ \bibinfo {address} {Singapore},\ \bibinfo {year} {2022})\ pp.\ \bibinfo {pages} {1--17},\ \Eprint {https://arxiv.org/abs/2207.03561} {arXiv:2207.03561 [astro-ph.HE]} \BibitemShut {NoStop}%
\bibitem [{\citenamefont {Johns}\ \emph {et~al.}(2025)\citenamefont {Johns}, \citenamefont {Richers},\ and\ \citenamefont {Wu}}]{Johns:2025mlm}%
  \BibitemOpen
  \bibfield  {author} {\bibinfo {author} {\bibfnamefont {L.}~\bibnamefont {Johns}}, \bibinfo {author} {\bibfnamefont {S.}~\bibnamefont {Richers}},\ and\ \bibinfo {author} {\bibfnamefont {M.-R.}\ \bibnamefont {Wu}},\ }\bibfield  {title} {\bibinfo {title} {{Neutrino Oscillations in Core-Collapse Supernovae and Neutron Star Mergers}},\ }\Eprint {https://arxiv.org/abs/2503.05959} {arXiv:2503.05959 [astro-ph.HE]}  (\bibinfo {year} {2025})\BibitemShut {NoStop}%
\bibitem [{\citenamefont {Stapleford}\ \emph {et~al.}(2020)\citenamefont {Stapleford}, \citenamefont {Fr\"ohlich},\ and\ \citenamefont {Kneller}}]{Stapleford:2019yqg}%
  \BibitemOpen
  \bibfield  {author} {\bibinfo {author} {\bibfnamefont {C.~J.}\ \bibnamefont {Stapleford}}, \bibinfo {author} {\bibfnamefont {C.}~\bibnamefont {Fr\"ohlich}},\ and\ \bibinfo {author} {\bibfnamefont {J.~P.}\ \bibnamefont {Kneller}},\ }\bibfield  {title} {\bibinfo {title} {{Coupling Neutrino Oscillations and Simulations of Core-Collapse Supernovae}},\ }\href {https://doi.org/10.1103/PhysRevD.102.081301} {\bibfield  {journal} {\bibinfo  {journal} {Phys. Rev. D}\ }\textbf {\bibinfo {volume} {102}},\ \bibinfo {pages} {081301} (\bibinfo {year} {2020})},\ \Eprint {https://arxiv.org/abs/1910.04172} {arXiv:1910.04172 [astro-ph.HE]} \BibitemShut {NoStop}%
\bibitem [{\citenamefont {Ehring}\ \emph {et~al.}(2023{\natexlab{a}})\citenamefont {Ehring}, \citenamefont {Abbar}, \citenamefont {Janka}, \citenamefont {Raffelt},\ and\ \citenamefont {Tamborra}}]{Ehring:2023lcd}%
  \BibitemOpen
  \bibfield  {author} {\bibinfo {author} {\bibfnamefont {J.}~\bibnamefont {Ehring}}, \bibinfo {author} {\bibfnamefont {S.}~\bibnamefont {Abbar}}, \bibinfo {author} {\bibfnamefont {H.-T.}\ \bibnamefont {Janka}}, \bibinfo {author} {\bibfnamefont {G.}~\bibnamefont {Raffelt}},\ and\ \bibinfo {author} {\bibfnamefont {I.}~\bibnamefont {Tamborra}},\ }\bibfield  {title} {\bibinfo {title} {{Fast neutrino flavor conversion in core-collapse supernovae: A parametric study in 1D models}},\ }\href {https://doi.org/10.1103/PhysRevD.107.103034} {\bibfield  {journal} {\bibinfo  {journal} {Phys. Rev. D}\ }\textbf {\bibinfo {volume} {107}},\ \bibinfo {pages} {103034} (\bibinfo {year} {2023}{\natexlab{a}})},\ \Eprint {https://arxiv.org/abs/2301.11938} {arXiv:2301.11938 [astro-ph.HE]} \BibitemShut {NoStop}%
\bibitem [{\citenamefont {Ehring}\ \emph {et~al.}(2023{\natexlab{b}})\citenamefont {Ehring}, \citenamefont {Abbar}, \citenamefont {Janka}, \citenamefont {Raffelt},\ and\ \citenamefont {Tamborra}}]{Ehring:2023abs}%
  \BibitemOpen
  \bibfield  {author} {\bibinfo {author} {\bibfnamefont {J.}~\bibnamefont {Ehring}}, \bibinfo {author} {\bibfnamefont {S.}~\bibnamefont {Abbar}}, \bibinfo {author} {\bibfnamefont {H.-T.}\ \bibnamefont {Janka}}, \bibinfo {author} {\bibfnamefont {G.}~\bibnamefont {Raffelt}},\ and\ \bibinfo {author} {\bibfnamefont {I.}~\bibnamefont {Tamborra}},\ }\bibfield  {title} {\bibinfo {title} {{Fast Neutrino Flavor Conversions Can Help and Hinder Neutrino-Driven Explosions}},\ }\href {https://doi.org/10.1103/PhysRevLett.131.061401} {\bibfield  {journal} {\bibinfo  {journal} {Phys. Rev. Lett.}\ }\textbf {\bibinfo {volume} {131}},\ \bibinfo {pages} {061401} (\bibinfo {year} {2023}{\natexlab{b}})},\ \Eprint {https://arxiv.org/abs/2305.11207} {arXiv:2305.11207 [astro-ph.HE]} \BibitemShut {NoStop}%
\bibitem [{\citenamefont {Mori}\ \emph {et~al.}(2025)\citenamefont {Mori}, \citenamefont {Takiwaki}, \citenamefont {Kotake},\ and\ \citenamefont {Horiuchi}}]{Mori:2025cke}%
  \BibitemOpen
  \bibfield  {author} {\bibinfo {author} {\bibfnamefont {K.}~\bibnamefont {Mori}}, \bibinfo {author} {\bibfnamefont {T.}~\bibnamefont {Takiwaki}}, \bibinfo {author} {\bibfnamefont {K.}~\bibnamefont {Kotake}},\ and\ \bibinfo {author} {\bibfnamefont {S.}~\bibnamefont {Horiuchi}},\ }\bibfield  {title} {\bibinfo {title} {{Three-dimensional core-collapse supernova models with phenomenological treatment of neutrino flavor conversions}},\ }\href {https://doi.org/10.1093/pasj/psaf007} {\bibfield  {journal} {\bibinfo  {journal} {Publ. Astron. Soc. Jap.}\ }\textbf {\bibinfo {volume} {77}},\ \bibinfo {pages} {L9} (\bibinfo {year} {2025})},\ \Eprint {https://arxiv.org/abs/2501.15256} {arXiv:2501.15256 [astro-ph.HE]} \BibitemShut {NoStop}%
\bibitem [{\citenamefont {Li}\ and\ \citenamefont {Siegel}(2021)}]{Li:2021vqj}%
  \BibitemOpen
  \bibfield  {author} {\bibinfo {author} {\bibfnamefont {X.}~\bibnamefont {Li}}\ and\ \bibinfo {author} {\bibfnamefont {D.~M.}\ \bibnamefont {Siegel}},\ }\bibfield  {title} {\bibinfo {title} {{Neutrino Fast Flavor Conversions in Neutron-Star Postmerger Accretion Disks}},\ }\href {https://doi.org/10.1103/PhysRevLett.126.251101} {\bibfield  {journal} {\bibinfo  {journal} {Phys. Rev. Lett.}\ }\textbf {\bibinfo {volume} {126}},\ \bibinfo {pages} {251101} (\bibinfo {year} {2021})},\ \Eprint {https://arxiv.org/abs/2103.02616} {arXiv:2103.02616 [astro-ph.HE]} \BibitemShut {NoStop}%
\bibitem [{\citenamefont {Just}\ \emph {et~al.}(2022)\citenamefont {Just}, \citenamefont {Abbar}, \citenamefont {Wu}, \citenamefont {Tamborra}, \citenamefont {Janka},\ and\ \citenamefont {Capozzi}}]{Just:2022flt}%
  \BibitemOpen
  \bibfield  {author} {\bibinfo {author} {\bibfnamefont {O.}~\bibnamefont {Just}}, \bibinfo {author} {\bibfnamefont {S.}~\bibnamefont {Abbar}}, \bibinfo {author} {\bibfnamefont {M.-R.}\ \bibnamefont {Wu}}, \bibinfo {author} {\bibfnamefont {I.}~\bibnamefont {Tamborra}}, \bibinfo {author} {\bibfnamefont {H.-T.}\ \bibnamefont {Janka}},\ and\ \bibinfo {author} {\bibfnamefont {F.}~\bibnamefont {Capozzi}},\ }\bibfield  {title} {\bibinfo {title} {{Fast neutrino conversion in hydrodynamic simulations of neutrino-cooled accretion disks}},\ }\href {https://doi.org/10.1103/PhysRevD.105.083024} {\bibfield  {journal} {\bibinfo  {journal} {Phys. Rev. D}\ }\textbf {\bibinfo {volume} {105}},\ \bibinfo {pages} {083024} (\bibinfo {year} {2022})},\ \Eprint {https://arxiv.org/abs/2203.16559} {arXiv:2203.16559 [astro-ph.HE]} \BibitemShut {NoStop}%
\bibitem [{\citenamefont {Fern\'andez}\ \emph {et~al.}(2022)\citenamefont {Fern\'andez}, \citenamefont {Richers}, \citenamefont {Mulyk},\ and\ \citenamefont {Fahlman}}]{Fernandez:2022yyv}%
  \BibitemOpen
  \bibfield  {author} {\bibinfo {author} {\bibfnamefont {R.}~\bibnamefont {Fern\'andez}}, \bibinfo {author} {\bibfnamefont {S.}~\bibnamefont {Richers}}, \bibinfo {author} {\bibfnamefont {N.}~\bibnamefont {Mulyk}},\ and\ \bibinfo {author} {\bibfnamefont {S.}~\bibnamefont {Fahlman}},\ }\bibfield  {title} {\bibinfo {title} {{Fast flavor instability in hypermassive neutron star disk outflows}},\ }\href {https://doi.org/10.1103/PhysRevD.106.103003} {\bibfield  {journal} {\bibinfo  {journal} {Phys. Rev. D}\ }\textbf {\bibinfo {volume} {106}},\ \bibinfo {pages} {103003} (\bibinfo {year} {2022})},\ \Eprint {https://arxiv.org/abs/2207.10680} {arXiv:2207.10680 [astro-ph.HE]} \BibitemShut {NoStop}%
\bibitem [{\citenamefont {Qiu}\ \emph {et~al.}(2025)\citenamefont {Qiu}, \citenamefont {Radice}, \citenamefont {Richers},\ and\ \citenamefont {Bhattacharyya}}]{Qiu:2025kgy}%
  \BibitemOpen
  \bibfield  {author} {\bibinfo {author} {\bibfnamefont {Y.}~\bibnamefont {Qiu}}, \bibinfo {author} {\bibfnamefont {D.}~\bibnamefont {Radice}}, \bibinfo {author} {\bibfnamefont {S.}~\bibnamefont {Richers}},\ and\ \bibinfo {author} {\bibfnamefont {M.}~\bibnamefont {Bhattacharyya}},\ }\bibfield  {title} {\bibinfo {title} {{Neutrino Flavor Transformation in Neutron Star Mergers}},\ }\Eprint {https://arxiv.org/abs/2503.11758} {arXiv:2503.11758 [astro-ph.HE]}  (\bibinfo {year} {2025})\BibitemShut {NoStop}%
\bibitem [{\citenamefont {Nagakura}\ and\ \citenamefont {Zaizen}(2022)}]{Nagakura:2022kic}%
  \BibitemOpen
  \bibfield  {author} {\bibinfo {author} {\bibfnamefont {H.}~\bibnamefont {Nagakura}}\ and\ \bibinfo {author} {\bibfnamefont {M.}~\bibnamefont {Zaizen}},\ }\bibfield  {title} {\bibinfo {title} {{Time-Dependent and Quasisteady Features of Fast Neutrino-Flavor Conversion}},\ }\href {https://doi.org/10.1103/PhysRevLett.129.261101} {\bibfield  {journal} {\bibinfo  {journal} {Phys. Rev. Lett.}\ }\textbf {\bibinfo {volume} {129}},\ \bibinfo {pages} {261101} (\bibinfo {year} {2022})},\ \Eprint {https://arxiv.org/abs/2206.04097} {arXiv:2206.04097 [astro-ph.HE]} \BibitemShut {NoStop}%
\bibitem [{\citenamefont {Xiong}\ \emph {et~al.}(2023{\natexlab{a}})\citenamefont {Xiong}, \citenamefont {Wu}, \citenamefont {Martinez-Pinedo}, \citenamefont {Fischer}, \citenamefont {George}, \citenamefont {Lin},\ and\ \citenamefont {Johns}}]{Xiong:2022vsy}%
  \BibitemOpen
  \bibfield  {author} {\bibinfo {author} {\bibfnamefont {Z.}~\bibnamefont {Xiong}}, \bibinfo {author} {\bibfnamefont {M.-R.}\ \bibnamefont {Wu}}, \bibinfo {author} {\bibfnamefont {G.}~\bibnamefont {Martinez-Pinedo}}, \bibinfo {author} {\bibfnamefont {T.}~\bibnamefont {Fischer}}, \bibinfo {author} {\bibfnamefont {M.}~\bibnamefont {George}}, \bibinfo {author} {\bibfnamefont {C.-Y.}\ \bibnamefont {Lin}},\ and\ \bibinfo {author} {\bibfnamefont {L.}~\bibnamefont {Johns}},\ }\bibfield  {title} {\bibinfo {title} {{Evolution of collisional neutrino flavor instabilities in spherically symmetric supernova models}},\ }\href {https://doi.org/10.1103/PhysRevD.107.083016} {\bibfield  {journal} {\bibinfo  {journal} {Phys. Rev. D}\ }\textbf {\bibinfo {volume} {107}},\ \bibinfo {pages} {083016} (\bibinfo {year} {2023}{\natexlab{a}})},\ \Eprint {https://arxiv.org/abs/2210.08254} {arXiv:2210.08254 [astro-ph.HE]} \BibitemShut {NoStop}%
\bibitem [{\citenamefont {Nagakura}\ and\ \citenamefont {Zaizen}(2023{\natexlab{a}})}]{Nagakura:2022xwe}%
  \BibitemOpen
  \bibfield  {author} {\bibinfo {author} {\bibfnamefont {H.}~\bibnamefont {Nagakura}}\ and\ \bibinfo {author} {\bibfnamefont {M.}~\bibnamefont {Zaizen}},\ }\bibfield  {title} {\bibinfo {title} {{Connecting small-scale to large-scale structures of fast neutrino-flavor conversion}},\ }\href {https://doi.org/10.1103/PhysRevD.107.063033} {\bibfield  {journal} {\bibinfo  {journal} {Phys. Rev. D}\ }\textbf {\bibinfo {volume} {107}},\ \bibinfo {pages} {063033} (\bibinfo {year} {2023}{\natexlab{a}})},\ \Eprint {https://arxiv.org/abs/2211.01398} {arXiv:2211.01398 [astro-ph.HE]} \BibitemShut {NoStop}%
\bibitem [{\citenamefont {Nagakura}(2023{\natexlab{a}})}]{Nagakura:2023mhr}%
  \BibitemOpen
  \bibfield  {author} {\bibinfo {author} {\bibfnamefont {H.}~\bibnamefont {Nagakura}},\ }\bibfield  {title} {\bibinfo {title} {{Roles of Fast Neutrino-Flavor Conversion on the Neutrino-Heating Mechanism of Core-Collapse Supernova}},\ }\href {https://doi.org/10.1103/PhysRevLett.130.211401} {\bibfield  {journal} {\bibinfo  {journal} {Phys. Rev. Lett.}\ }\textbf {\bibinfo {volume} {130}},\ \bibinfo {pages} {211401} (\bibinfo {year} {2023}{\natexlab{a}})},\ \Eprint {https://arxiv.org/abs/2301.10785} {arXiv:2301.10785 [astro-ph.HE]} \BibitemShut {NoStop}%
\bibitem [{\citenamefont {Nagakura}(2023{\natexlab{b}})}]{Nagakura:2023wbf}%
  \BibitemOpen
  \bibfield  {author} {\bibinfo {author} {\bibfnamefont {H.}~\bibnamefont {Nagakura}},\ }\bibfield  {title} {\bibinfo {title} {{Global features of fast neutrino-flavor conversion in binary neutron star mergers}},\ }\href {https://doi.org/10.1103/PhysRevD.108.103014} {\bibfield  {journal} {\bibinfo  {journal} {Phys. Rev. D}\ }\textbf {\bibinfo {volume} {108}},\ \bibinfo {pages} {103014} (\bibinfo {year} {2023}{\natexlab{b}})},\ \Eprint {https://arxiv.org/abs/2306.10108} {arXiv:2306.10108 [astro-ph.HE]} \BibitemShut {NoStop}%
\bibitem [{\citenamefont {Shalgar}\ and\ \citenamefont {Tamborra}(2024{\natexlab{a}})}]{Shalgar:2024gjt}%
  \BibitemOpen
  \bibfield  {author} {\bibinfo {author} {\bibfnamefont {S.}~\bibnamefont {Shalgar}}\ and\ \bibinfo {author} {\bibfnamefont {I.}~\bibnamefont {Tamborra}},\ }\bibfield  {title} {\bibinfo {title} {{Neutrino quantum kinetics in a core-collapse supernova}},\ }\href {https://doi.org/10.1088/1475-7516/2024/09/021} {\bibfield  {journal} {\bibinfo  {journal} {JCAP}\ }\textbf {\bibinfo {volume} {09}},\ \bibinfo {pages} {021}},\ \Eprint {https://arxiv.org/abs/2406.09504} {arXiv:2406.09504 [astro-ph.HE]} \BibitemShut {NoStop}%
\bibitem [{\citenamefont {Myers}\ \emph {et~al.}(2022)\citenamefont {Myers}, \citenamefont {Cooper}, \citenamefont {Warren}, \citenamefont {Kneller}, \citenamefont {McLaughlin}, \citenamefont {Richers}, \citenamefont {Grohs},\ and\ \citenamefont {Frohlich}}]{Myers:2021hnp}%
  \BibitemOpen
  \bibfield  {author} {\bibinfo {author} {\bibfnamefont {M.}~\bibnamefont {Myers}}, \bibinfo {author} {\bibfnamefont {T.}~\bibnamefont {Cooper}}, \bibinfo {author} {\bibfnamefont {M.}~\bibnamefont {Warren}}, \bibinfo {author} {\bibfnamefont {J.}~\bibnamefont {Kneller}}, \bibinfo {author} {\bibfnamefont {G.}~\bibnamefont {McLaughlin}}, \bibinfo {author} {\bibfnamefont {S.}~\bibnamefont {Richers}}, \bibinfo {author} {\bibfnamefont {E.}~\bibnamefont {Grohs}},\ and\ \bibinfo {author} {\bibfnamefont {C.}~\bibnamefont {Frohlich}},\ }\bibfield  {title} {\bibinfo {title} {{Neutrino flavor mixing with moments}},\ }\href {https://doi.org/10.1103/PhysRevD.105.123036} {\bibfield  {journal} {\bibinfo  {journal} {Phys. Rev. D}\ }\textbf {\bibinfo {volume} {105}},\ \bibinfo {pages} {123036} (\bibinfo {year} {2022})},\ \Eprint {https://arxiv.org/abs/2111.13722} {arXiv:2111.13722 [hep-ph]} \BibitemShut {NoStop}%
\bibitem [{\citenamefont {Grohs}\ \emph {et~al.}(2023)\citenamefont {Grohs}, \citenamefont {Richers}, \citenamefont {Couch}, \citenamefont {Foucart}, \citenamefont {Kneller},\ and\ \citenamefont {McLaughlin}}]{Grohs:2022fyq}%
  \BibitemOpen
  \bibfield  {author} {\bibinfo {author} {\bibfnamefont {E.}~\bibnamefont {Grohs}}, \bibinfo {author} {\bibfnamefont {S.}~\bibnamefont {Richers}}, \bibinfo {author} {\bibfnamefont {S.~M.}\ \bibnamefont {Couch}}, \bibinfo {author} {\bibfnamefont {F.}~\bibnamefont {Foucart}}, \bibinfo {author} {\bibfnamefont {J.~P.}\ \bibnamefont {Kneller}},\ and\ \bibinfo {author} {\bibfnamefont {G.~C.}\ \bibnamefont {McLaughlin}},\ }\bibfield  {title} {\bibinfo {title} {{Neutrino fast flavor instability in three dimensions for a neutron star merger}},\ }\href {https://doi.org/10.1016/j.physletb.2023.138210} {\bibfield  {journal} {\bibinfo  {journal} {Phys. Lett. B}\ }\textbf {\bibinfo {volume} {846}},\ \bibinfo {pages} {138210} (\bibinfo {year} {2023})},\ \Eprint {https://arxiv.org/abs/2207.02214} {arXiv:2207.02214 [hep-ph]} \BibitemShut {NoStop}%
\bibitem [{\citenamefont {Grohs}\ \emph {et~al.}(2024)\citenamefont {Grohs}, \citenamefont {Richers}, \citenamefont {Couch}, \citenamefont {Foucart}, \citenamefont {Froustey}, \citenamefont {Kneller},\ and\ \citenamefont {McLaughlin}}]{Grohs:2023pgq}%
  \BibitemOpen
  \bibfield  {author} {\bibinfo {author} {\bibfnamefont {E.}~\bibnamefont {Grohs}}, \bibinfo {author} {\bibfnamefont {S.}~\bibnamefont {Richers}}, \bibinfo {author} {\bibfnamefont {S.~M.}\ \bibnamefont {Couch}}, \bibinfo {author} {\bibfnamefont {F.}~\bibnamefont {Foucart}}, \bibinfo {author} {\bibfnamefont {J.}~\bibnamefont {Froustey}}, \bibinfo {author} {\bibfnamefont {J.~P.}\ \bibnamefont {Kneller}},\ and\ \bibinfo {author} {\bibfnamefont {G.~C.}\ \bibnamefont {McLaughlin}},\ }\bibfield  {title} {\bibinfo {title} {{Two-moment Neutrino Flavor Transformation with Applications to the Fast Flavor Instability in Neutron Star Mergers}},\ }\href {https://doi.org/10.3847/1538-4357/ad13f2} {\bibfield  {journal} {\bibinfo  {journal} {Astrophys. J.}\ }\textbf {\bibinfo {volume} {963}},\ \bibinfo {pages} {11} (\bibinfo {year} {2024})},\ \Eprint {https://arxiv.org/abs/2309.00972} {arXiv:2309.00972 [astro-ph.HE]} \BibitemShut {NoStop}%
\bibitem [{\citenamefont {Froustey}\ \emph {et~al.}(2024)\citenamefont {Froustey}, \citenamefont {Richers}, \citenamefont {Grohs}, \citenamefont {Flynn}, \citenamefont {Foucart}, \citenamefont {Kneller},\ and\ \citenamefont {McLaughlin}}]{Froustey:2023skf}%
  \BibitemOpen
  \bibfield  {author} {\bibinfo {author} {\bibfnamefont {J.}~\bibnamefont {Froustey}}, \bibinfo {author} {\bibfnamefont {S.}~\bibnamefont {Richers}}, \bibinfo {author} {\bibfnamefont {E.}~\bibnamefont {Grohs}}, \bibinfo {author} {\bibfnamefont {S.~D.}\ \bibnamefont {Flynn}}, \bibinfo {author} {\bibfnamefont {F.}~\bibnamefont {Foucart}}, \bibinfo {author} {\bibfnamefont {J.~P.}\ \bibnamefont {Kneller}},\ and\ \bibinfo {author} {\bibfnamefont {G.~C.}\ \bibnamefont {McLaughlin}},\ }\bibfield  {title} {\bibinfo {title} {{Neutrino fast flavor oscillations with moments: Linear stability analysis and application to neutron star mergers}},\ }\href {https://doi.org/10.1103/PhysRevD.109.043046} {\bibfield  {journal} {\bibinfo  {journal} {Phys. Rev. D}\ }\textbf {\bibinfo {volume} {109}},\ \bibinfo {pages} {043046} (\bibinfo {year} {2024})},\ \Eprint {https://arxiv.org/abs/2311.11968} {arXiv:2311.11968 [astro-ph.HE]} \BibitemShut {NoStop}%
\bibitem [{\citenamefont {Froustey}\ \emph {et~al.}(2025)\citenamefont {Froustey}, \citenamefont {Kneller},\ and\ \citenamefont {McLaughlin}}]{Froustey:2024sgz}%
  \BibitemOpen
  \bibfield  {author} {\bibinfo {author} {\bibfnamefont {J.}~\bibnamefont {Froustey}}, \bibinfo {author} {\bibfnamefont {J.~P.}\ \bibnamefont {Kneller}},\ and\ \bibinfo {author} {\bibfnamefont {G.~C.}\ \bibnamefont {McLaughlin}},\ }\bibfield  {title} {\bibinfo {title} {{Quantum maximum entropy closure for small flavor coherence}},\ }\href {https://doi.org/10.1103/PhysRevD.111.063022} {\bibfield  {journal} {\bibinfo  {journal} {Phys. Rev. D}\ }\textbf {\bibinfo {volume} {111}},\ \bibinfo {pages} {063022} (\bibinfo {year} {2025})},\ \Eprint {https://arxiv.org/abs/2409.05807} {arXiv:2409.05807 [hep-ph]} \BibitemShut {NoStop}%
\bibitem [{\citenamefont {Kneller}\ \emph {et~al.}(2025)\citenamefont {Kneller}, \citenamefont {Froustey}, \citenamefont {Grohs}, \citenamefont {Foucart}, \citenamefont {McLaughlin},\ and\ \citenamefont {Richers}}]{Kneller:2024buy}%
  \BibitemOpen
  \bibfield  {author} {\bibinfo {author} {\bibfnamefont {J.~P.}\ \bibnamefont {Kneller}}, \bibinfo {author} {\bibfnamefont {J.}~\bibnamefont {Froustey}}, \bibinfo {author} {\bibfnamefont {E.~B.}\ \bibnamefont {Grohs}}, \bibinfo {author} {\bibfnamefont {F.}~\bibnamefont {Foucart}}, \bibinfo {author} {\bibfnamefont {G.~C.}\ \bibnamefont {McLaughlin}},\ and\ \bibinfo {author} {\bibfnamefont {S.}~\bibnamefont {Richers}},\ }\bibfield  {title} {\bibinfo {title} {{Quantum closures for neutrino moment transport}},\ }\href {https://doi.org/10.1103/PhysRevD.111.063046} {\bibfield  {journal} {\bibinfo  {journal} {Phys. Rev. D}\ }\textbf {\bibinfo {volume} {111}},\ \bibinfo {pages} {063046} (\bibinfo {year} {2025})},\ \Eprint {https://arxiv.org/abs/2410.00719} {arXiv:2410.00719 [hep-ph]} \BibitemShut {NoStop}%
\bibitem [{\citenamefont {Grohs}\ \emph {et~al.}(2025)\citenamefont {Grohs}, \citenamefont {Richers}, \citenamefont {Froustey}, \citenamefont {Foucart}, \citenamefont {Kneller},\ and\ \citenamefont {McLaughlin}}]{Grohs:2025ajr}%
  \BibitemOpen
  \bibfield  {author} {\bibinfo {author} {\bibfnamefont {E.}~\bibnamefont {Grohs}}, \bibinfo {author} {\bibfnamefont {S.}~\bibnamefont {Richers}}, \bibinfo {author} {\bibfnamefont {J.}~\bibnamefont {Froustey}}, \bibinfo {author} {\bibfnamefont {F.}~\bibnamefont {Foucart}}, \bibinfo {author} {\bibfnamefont {J.~P.}\ \bibnamefont {Kneller}},\ and\ \bibinfo {author} {\bibfnamefont {G.~C.}\ \bibnamefont {McLaughlin}},\ }\bibfield  {title} {\bibinfo {title} {{Advection algorithms for quantum neutrino moment transport}},\ }\href {https://doi.org/10.1103/PhysRevD.111.083018} {\bibfield  {journal} {\bibinfo  {journal} {Phys. Rev. D}\ }\textbf {\bibinfo {volume} {111}},\ \bibinfo {pages} {083018} (\bibinfo {year} {2025})},\ \Eprint {https://arxiv.org/abs/2501.07540} {arXiv:2501.07540 [astro-ph.HE]} \BibitemShut {NoStop}%
\bibitem [{\citenamefont {Nagakura}\ \emph {et~al.}(2024)\citenamefont {Nagakura}, \citenamefont {Johns},\ and\ \citenamefont {Zaizen}}]{Nagakura:2023jfi}%
  \BibitemOpen
  \bibfield  {author} {\bibinfo {author} {\bibfnamefont {H.}~\bibnamefont {Nagakura}}, \bibinfo {author} {\bibfnamefont {L.}~\bibnamefont {Johns}},\ and\ \bibinfo {author} {\bibfnamefont {M.}~\bibnamefont {Zaizen}},\ }\bibfield  {title} {\bibinfo {title} {{Bhatnagar-Gross-Krook subgrid model for neutrino quantum kinetics}},\ }\href {https://doi.org/10.1103/PhysRevD.109.083013} {\bibfield  {journal} {\bibinfo  {journal} {Phys. Rev. D}\ }\textbf {\bibinfo {volume} {109}},\ \bibinfo {pages} {083013} (\bibinfo {year} {2024})},\ \Eprint {https://arxiv.org/abs/2312.16285} {arXiv:2312.16285 [astro-ph.HE]} \BibitemShut {NoStop}%
\bibitem [{\citenamefont {Bhattacharyya}\ and\ \citenamefont {Dasgupta}(2021)}]{Bhattacharyya:2020jpj}%
  \BibitemOpen
  \bibfield  {author} {\bibinfo {author} {\bibfnamefont {S.}~\bibnamefont {Bhattacharyya}}\ and\ \bibinfo {author} {\bibfnamefont {B.}~\bibnamefont {Dasgupta}},\ }\bibfield  {title} {\bibinfo {title} {{Fast Flavor Depolarization of Supernova Neutrinos}},\ }\href {https://doi.org/10.1103/PhysRevLett.126.061302} {\bibfield  {journal} {\bibinfo  {journal} {Phys. Rev. Lett.}\ }\textbf {\bibinfo {volume} {126}},\ \bibinfo {pages} {061302} (\bibinfo {year} {2021})},\ \Eprint {https://arxiv.org/abs/2009.03337} {arXiv:2009.03337 [hep-ph]} \BibitemShut {NoStop}%
\bibitem [{\citenamefont {Bhattacharyya}\ and\ \citenamefont {Dasgupta}(2022)}]{Bhattacharyya:2022eed}%
  \BibitemOpen
  \bibfield  {author} {\bibinfo {author} {\bibfnamefont {S.}~\bibnamefont {Bhattacharyya}}\ and\ \bibinfo {author} {\bibfnamefont {B.}~\bibnamefont {Dasgupta}},\ }\bibfield  {title} {\bibinfo {title} {{Elaborating the ultimate fate of fast collective neutrino flavor oscillations}},\ }\href {https://doi.org/10.1103/PhysRevD.106.103039} {\bibfield  {journal} {\bibinfo  {journal} {Phys. Rev. D}\ }\textbf {\bibinfo {volume} {106}},\ \bibinfo {pages} {103039} (\bibinfo {year} {2022})},\ \Eprint {https://arxiv.org/abs/2205.05129} {arXiv:2205.05129 [hep-ph]} \BibitemShut {NoStop}%
\bibitem [{\citenamefont {Zaizen}\ and\ \citenamefont {Nagakura}(2023{\natexlab{a}})}]{Zaizen:2022cik}%
  \BibitemOpen
  \bibfield  {author} {\bibinfo {author} {\bibfnamefont {M.}~\bibnamefont {Zaizen}}\ and\ \bibinfo {author} {\bibfnamefont {H.}~\bibnamefont {Nagakura}},\ }\bibfield  {title} {\bibinfo {title} {{Simple method for determining asymptotic states of fast neutrino-flavor conversion}},\ }\href {https://doi.org/10.1103/PhysRevD.107.103022} {\bibfield  {journal} {\bibinfo  {journal} {Phys. Rev. D}\ }\textbf {\bibinfo {volume} {107}},\ \bibinfo {pages} {103022} (\bibinfo {year} {2023}{\natexlab{a}})},\ \Eprint {https://arxiv.org/abs/2211.09343} {arXiv:2211.09343 [astro-ph.HE]} \BibitemShut {NoStop}%
\bibitem [{\citenamefont {Zaizen}\ and\ \citenamefont {Nagakura}(2023{\natexlab{b}})}]{Zaizen:2023ihz}%
  \BibitemOpen
  \bibfield  {author} {\bibinfo {author} {\bibfnamefont {M.}~\bibnamefont {Zaizen}}\ and\ \bibinfo {author} {\bibfnamefont {H.}~\bibnamefont {Nagakura}},\ }\bibfield  {title} {\bibinfo {title} {{Characterizing quasisteady states of fast neutrino-flavor conversion by stability and conservation laws}},\ }\href {https://doi.org/10.1103/PhysRevD.107.123021} {\bibfield  {journal} {\bibinfo  {journal} {Phys. Rev. D}\ }\textbf {\bibinfo {volume} {107}},\ \bibinfo {pages} {123021} (\bibinfo {year} {2023}{\natexlab{b}})},\ \Eprint {https://arxiv.org/abs/2304.05044} {arXiv:2304.05044 [astro-ph.HE]} \BibitemShut {NoStop}%
\bibitem [{\citenamefont {Nagakura}\ and\ \citenamefont {Zaizen}(2023{\natexlab{b}})}]{Nagakura:2023xhc}%
  \BibitemOpen
  \bibfield  {author} {\bibinfo {author} {\bibfnamefont {H.}~\bibnamefont {Nagakura}}\ and\ \bibinfo {author} {\bibfnamefont {M.}~\bibnamefont {Zaizen}},\ }\bibfield  {title} {\bibinfo {title} {{Basic characteristics of neutrino flavor conversions in the postshock regions of core-collapse supernova}},\ }\href {https://doi.org/10.1103/PhysRevD.108.123003} {\bibfield  {journal} {\bibinfo  {journal} {Phys. Rev. D}\ }\textbf {\bibinfo {volume} {108}},\ \bibinfo {pages} {123003} (\bibinfo {year} {2023}{\natexlab{b}})},\ \Eprint {https://arxiv.org/abs/2308.14800} {arXiv:2308.14800 [astro-ph.HE]} \BibitemShut {NoStop}%
\bibitem [{\citenamefont {Abbar}\ \emph {et~al.}(2024)\citenamefont {Abbar}, \citenamefont {Wu},\ and\ \citenamefont {Xiong}}]{Abbar:2023ltx}%
  \BibitemOpen
  \bibfield  {author} {\bibinfo {author} {\bibfnamefont {S.}~\bibnamefont {Abbar}}, \bibinfo {author} {\bibfnamefont {M.-R.}\ \bibnamefont {Wu}},\ and\ \bibinfo {author} {\bibfnamefont {Z.}~\bibnamefont {Xiong}},\ }\bibfield  {title} {\bibinfo {title} {{Physics-informed neural networks for predicting the asymptotic outcome of fast neutrino flavor conversions}},\ }\href {https://doi.org/10.1103/PhysRevD.109.043024} {\bibfield  {journal} {\bibinfo  {journal} {Phys. Rev. D}\ }\textbf {\bibinfo {volume} {109}},\ \bibinfo {pages} {043024} (\bibinfo {year} {2024})},\ \Eprint {https://arxiv.org/abs/2311.15656} {arXiv:2311.15656 [astro-ph.HE]} \BibitemShut {NoStop}%
\bibitem [{\citenamefont {Xiong}\ \emph {et~al.}(2025)\citenamefont {Xiong}, \citenamefont {Wu}, \citenamefont {George},\ and\ \citenamefont {Lin}}]{Xiong:2024pue}%
  \BibitemOpen
  \bibfield  {author} {\bibinfo {author} {\bibfnamefont {Z.}~\bibnamefont {Xiong}}, \bibinfo {author} {\bibfnamefont {M.-R.}\ \bibnamefont {Wu}}, \bibinfo {author} {\bibfnamefont {M.}~\bibnamefont {George}},\ and\ \bibinfo {author} {\bibfnamefont {C.-Y.}\ \bibnamefont {Lin}},\ }\bibfield  {title} {\bibinfo {title} {{Robust Integration of Fast Flavor Conversions in Classical Neutrino Transport}},\ }\href {https://doi.org/10.1103/PhysRevLett.134.051003} {\bibfield  {journal} {\bibinfo  {journal} {Phys. Rev. Lett.}\ }\textbf {\bibinfo {volume} {134}},\ \bibinfo {pages} {051003} (\bibinfo {year} {2025})},\ \Eprint {https://arxiv.org/abs/2403.17269} {arXiv:2403.17269 [astro-ph.HE]} \BibitemShut {NoStop}%
\bibitem [{\citenamefont {Richers}\ \emph {et~al.}(2024)\citenamefont {Richers}, \citenamefont {Froustey}, \citenamefont {Ghosh}, \citenamefont {Foucart},\ and\ \citenamefont {Gomez}}]{Richers:2024zit}%
  \BibitemOpen
  \bibfield  {author} {\bibinfo {author} {\bibfnamefont {S.}~\bibnamefont {Richers}}, \bibinfo {author} {\bibfnamefont {J.}~\bibnamefont {Froustey}}, \bibinfo {author} {\bibfnamefont {S.}~\bibnamefont {Ghosh}}, \bibinfo {author} {\bibfnamefont {F.}~\bibnamefont {Foucart}},\ and\ \bibinfo {author} {\bibfnamefont {J.}~\bibnamefont {Gomez}},\ }\bibfield  {title} {\bibinfo {title} {{Asymptotic-state prediction for fast flavor transformation in neutron star mergers}},\ }\href {https://doi.org/10.1103/PhysRevD.110.103019} {\bibfield  {journal} {\bibinfo  {journal} {Phys. Rev. D}\ }\textbf {\bibinfo {volume} {110}},\ \bibinfo {pages} {103019} (\bibinfo {year} {2024})},\ \Eprint {https://arxiv.org/abs/2409.04405} {arXiv:2409.04405 [astro-ph.HE]} \BibitemShut {NoStop}%
\bibitem [{\citenamefont {George}\ \emph {et~al.}(2024)\citenamefont {George}, \citenamefont {Xiong}, \citenamefont {Wu},\ and\ \citenamefont {Lin}}]{George:2024zxz}%
  \BibitemOpen
  \bibfield  {author} {\bibinfo {author} {\bibfnamefont {M.}~\bibnamefont {George}}, \bibinfo {author} {\bibfnamefont {Z.}~\bibnamefont {Xiong}}, \bibinfo {author} {\bibfnamefont {M.-R.}\ \bibnamefont {Wu}},\ and\ \bibinfo {author} {\bibfnamefont {C.-Y.}\ \bibnamefont {Lin}},\ }\bibfield  {title} {\bibinfo {title} {{Evolution and the quasistationary state of collective fast neutrino flavor conversion in three dimensions without axisymmetry}},\ }\href {https://doi.org/10.1103/PhysRevD.110.123018} {\bibfield  {journal} {\bibinfo  {journal} {Phys. Rev. D}\ }\textbf {\bibinfo {volume} {110}},\ \bibinfo {pages} {123018} (\bibinfo {year} {2024})},\ \Eprint {https://arxiv.org/abs/2409.08833} {arXiv:2409.08833 [astro-ph.HE]} \BibitemShut {NoStop}%
\bibitem [{\citenamefont {Lund}\ \emph {et~al.}(2025)\citenamefont {Lund}, \citenamefont {Mukhopadhyay}, \citenamefont {Miller},\ and\ \citenamefont {McLaughlin}}]{Lund:2025jjo}%
  \BibitemOpen
  \bibfield  {author} {\bibinfo {author} {\bibfnamefont {K.~A.}\ \bibnamefont {Lund}}, \bibinfo {author} {\bibfnamefont {P.}~\bibnamefont {Mukhopadhyay}}, \bibinfo {author} {\bibfnamefont {J.~M.}\ \bibnamefont {Miller}},\ and\ \bibinfo {author} {\bibfnamefont {G.~C.}\ \bibnamefont {McLaughlin}},\ }\bibfield  {title} {\bibinfo {title} {{Angle-dependent in Situ Fast Flavor Transformations in Post-neutron-star-merger Disks}},\ }\href {https://doi.org/10.3847/2041-8213/add0a7} {\bibfield  {journal} {\bibinfo  {journal} {Astrophys. J. Lett.}\ }\textbf {\bibinfo {volume} {985}},\ \bibinfo {pages} {L9} (\bibinfo {year} {2025})},\ \Eprint {https://arxiv.org/abs/2503.23727} {arXiv:2503.23727 [astro-ph.HE]} \BibitemShut {NoStop}%
\bibitem [{\citenamefont {Wang}\ and\ \citenamefont {Burrows}(2025)}]{Wang:2025nii}%
  \BibitemOpen
  \bibfield  {author} {\bibinfo {author} {\bibfnamefont {T.}~\bibnamefont {Wang}}\ and\ \bibinfo {author} {\bibfnamefont {A.}~\bibnamefont {Burrows}},\ }\bibfield  {title} {\bibinfo {title} {{The Effect of the Fast-Flavor Instability on Core-Collapse Supernova Models}},\ }\Eprint {https://arxiv.org/abs/2503.04896} {arXiv:2503.04896 [astro-ph.HE]}  (\bibinfo {year} {2025})\BibitemShut {NoStop}%
\bibitem [{\citenamefont {Johns}(2023{\natexlab{a}})}]{Johns:2021qby}%
  \BibitemOpen
  \bibfield  {author} {\bibinfo {author} {\bibfnamefont {L.}~\bibnamefont {Johns}},\ }\bibfield  {title} {\bibinfo {title} {{Collisional Flavor Instabilities of Supernova Neutrinos}},\ }\href {https://doi.org/10.1103/PhysRevLett.130.191001} {\bibfield  {journal} {\bibinfo  {journal} {Phys. Rev. Lett.}\ }\textbf {\bibinfo {volume} {130}},\ \bibinfo {pages} {191001} (\bibinfo {year} {2023}{\natexlab{a}})},\ \Eprint {https://arxiv.org/abs/2104.11369} {arXiv:2104.11369 [hep-ph]} \BibitemShut {NoStop}%
\bibitem [{\citenamefont {Xiong}\ \emph {et~al.}(2023{\natexlab{b}})\citenamefont {Xiong}, \citenamefont {Johns}, \citenamefont {Wu},\ and\ \citenamefont {Duan}}]{Xiong:2022zqz}%
  \BibitemOpen
  \bibfield  {author} {\bibinfo {author} {\bibfnamefont {Z.}~\bibnamefont {Xiong}}, \bibinfo {author} {\bibfnamefont {L.}~\bibnamefont {Johns}}, \bibinfo {author} {\bibfnamefont {M.-R.}\ \bibnamefont {Wu}},\ and\ \bibinfo {author} {\bibfnamefont {H.}~\bibnamefont {Duan}},\ }\bibfield  {title} {\bibinfo {title} {{Collisional flavor instability in dense neutrino gases}},\ }\href {https://doi.org/10.1103/PhysRevD.108.083002} {\bibfield  {journal} {\bibinfo  {journal} {Phys. Rev. D}\ }\textbf {\bibinfo {volume} {108}},\ \bibinfo {pages} {083002} (\bibinfo {year} {2023}{\natexlab{b}})},\ \Eprint {https://arxiv.org/abs/2212.03750} {arXiv:2212.03750 [hep-ph]} \BibitemShut {NoStop}%
\bibitem [{\citenamefont {Lin}\ and\ \citenamefont {Duan}(2023)}]{Lin:2022dek}%
  \BibitemOpen
  \bibfield  {author} {\bibinfo {author} {\bibfnamefont {Y.-C.}\ \bibnamefont {Lin}}\ and\ \bibinfo {author} {\bibfnamefont {H.}~\bibnamefont {Duan}},\ }\bibfield  {title} {\bibinfo {title} {{Collision-induced flavor instability in dense neutrino gases with energy-dependent scattering}},\ }\href {https://doi.org/10.1103/PhysRevD.107.083034} {\bibfield  {journal} {\bibinfo  {journal} {Phys. Rev. D}\ }\textbf {\bibinfo {volume} {107}},\ \bibinfo {pages} {083034} (\bibinfo {year} {2023})},\ \Eprint {https://arxiv.org/abs/2210.09218} {arXiv:2210.09218 [hep-ph]} \BibitemShut {NoStop}%
\bibitem [{\citenamefont {Padilla-Gay}\ \emph {et~al.}(2022)\citenamefont {Padilla-Gay}, \citenamefont {Tamborra},\ and\ \citenamefont {Raffelt}}]{Padilla-Gay:2022wck}%
  \BibitemOpen
  \bibfield  {author} {\bibinfo {author} {\bibfnamefont {I.}~\bibnamefont {Padilla-Gay}}, \bibinfo {author} {\bibfnamefont {I.}~\bibnamefont {Tamborra}},\ and\ \bibinfo {author} {\bibfnamefont {G.~G.}\ \bibnamefont {Raffelt}},\ }\bibfield  {title} {\bibinfo {title} {{Neutrino fast flavor pendulum. II. Collisional damping}},\ }\href {https://doi.org/10.1103/PhysRevD.106.103031} {\bibfield  {journal} {\bibinfo  {journal} {Phys. Rev. D}\ }\textbf {\bibinfo {volume} {106}},\ \bibinfo {pages} {103031} (\bibinfo {year} {2022})},\ \Eprint {https://arxiv.org/abs/2209.11235} {arXiv:2209.11235 [hep-ph]} \BibitemShut {NoStop}%
\bibitem [{\citenamefont {Liu}\ \emph {et~al.}(2023{\natexlab{a}})\citenamefont {Liu}, \citenamefont {Zaizen},\ and\ \citenamefont {Yamada}}]{Liu:2023pjw}%
  \BibitemOpen
  \bibfield  {author} {\bibinfo {author} {\bibfnamefont {J.}~\bibnamefont {Liu}}, \bibinfo {author} {\bibfnamefont {M.}~\bibnamefont {Zaizen}},\ and\ \bibinfo {author} {\bibfnamefont {S.}~\bibnamefont {Yamada}},\ }\bibfield  {title} {\bibinfo {title} {{Systematic study of the resonancelike structure in the collisional flavor instability of neutrinos}},\ }\href {https://doi.org/10.1103/PhysRevD.107.123011} {\bibfield  {journal} {\bibinfo  {journal} {Phys. Rev. D}\ }\textbf {\bibinfo {volume} {107}},\ \bibinfo {pages} {123011} (\bibinfo {year} {2023}{\natexlab{a}})},\ \Eprint {https://arxiv.org/abs/2302.06263} {arXiv:2302.06263 [hep-ph]} \BibitemShut {NoStop}%
\bibitem [{\citenamefont {Zaizen}\ \emph {et~al.}(2024)\citenamefont {Zaizen}, \citenamefont {Richers}, \citenamefont {Nagakura}, \citenamefont {Suzuki},\ and\ \citenamefont {Kato}}]{Zaizen:2024faj}%
  \BibitemOpen
  \bibfield  {author} {\bibinfo {author} {\bibfnamefont {M.}~\bibnamefont {Zaizen}}, \bibinfo {author} {\bibfnamefont {S.}~\bibnamefont {Richers}}, \bibinfo {author} {\bibfnamefont {H.}~\bibnamefont {Nagakura}}, \bibinfo {author} {\bibfnamefont {H.}~\bibnamefont {Suzuki}},\ and\ \bibinfo {author} {\bibfnamefont {C.}~\bibnamefont {Kato}},\ }\bibfield  {title} {\bibinfo {title} {{Inspecting neutrino flavor instabilities during proto-neutron star cooling phase in supernova: I. Spherically symmetric model}},\ }\Eprint {https://arxiv.org/abs/2407.20548} {arXiv:2407.20548 [astro-ph.HE]}  (\bibinfo {year} {2024})\BibitemShut {NoStop}%
\bibitem [{\citenamefont {Shalgar}\ and\ \citenamefont {Tamborra}(2024{\natexlab{b}})}]{Shalgar:2023aca}%
  \BibitemOpen
  \bibfield  {author} {\bibinfo {author} {\bibfnamefont {S.}~\bibnamefont {Shalgar}}\ and\ \bibinfo {author} {\bibfnamefont {I.}~\bibnamefont {Tamborra}},\ }\bibfield  {title} {\bibinfo {title} {{Do neutrinos become flavor unstable due to collisions with matter in the supernova decoupling region?}},\ }\href {https://doi.org/10.1103/PhysRevD.109.103011} {\bibfield  {journal} {\bibinfo  {journal} {Phys. Rev. D}\ }\textbf {\bibinfo {volume} {109}},\ \bibinfo {pages} {103011} (\bibinfo {year} {2024}{\natexlab{b}})},\ \Eprint {https://arxiv.org/abs/2307.10366} {arXiv:2307.10366 [astro-ph.HE]} \BibitemShut {NoStop}%
\bibitem [{\citenamefont {Liu}\ \emph {et~al.}(2023{\natexlab{b}})\citenamefont {Liu}, \citenamefont {Nagakura}, \citenamefont {Akaho}, \citenamefont {Ito}, \citenamefont {Zaizen},\ and\ \citenamefont {Yamada}}]{Liu:2023vtz}%
  \BibitemOpen
  \bibfield  {author} {\bibinfo {author} {\bibfnamefont {J.}~\bibnamefont {Liu}}, \bibinfo {author} {\bibfnamefont {H.}~\bibnamefont {Nagakura}}, \bibinfo {author} {\bibfnamefont {R.}~\bibnamefont {Akaho}}, \bibinfo {author} {\bibfnamefont {A.}~\bibnamefont {Ito}}, \bibinfo {author} {\bibfnamefont {M.}~\bibnamefont {Zaizen}},\ and\ \bibinfo {author} {\bibfnamefont {S.}~\bibnamefont {Yamada}},\ }\bibfield  {title} {\bibinfo {title} {{Universality of the neutrino collisional flavor instability in core-collapse supernovae}},\ }\href {https://doi.org/10.1103/PhysRevD.108.123024} {\bibfield  {journal} {\bibinfo  {journal} {Phys. Rev. D}\ }\textbf {\bibinfo {volume} {108}},\ \bibinfo {pages} {123024} (\bibinfo {year} {2023}{\natexlab{b}})},\ \Eprint {https://arxiv.org/abs/2310.05050} {arXiv:2310.05050 [astro-ph.HE]} \BibitemShut {NoStop}%
\bibitem [{\citenamefont {Akaho}\ \emph {et~al.}(2024)\citenamefont {Akaho}, \citenamefont {Liu}, \citenamefont {Nagakura}, \citenamefont {Zaizen},\ and\ \citenamefont {Yamada}}]{Akaho:2023brj}%
  \BibitemOpen
  \bibfield  {author} {\bibinfo {author} {\bibfnamefont {R.}~\bibnamefont {Akaho}}, \bibinfo {author} {\bibfnamefont {J.}~\bibnamefont {Liu}}, \bibinfo {author} {\bibfnamefont {H.}~\bibnamefont {Nagakura}}, \bibinfo {author} {\bibfnamefont {M.}~\bibnamefont {Zaizen}},\ and\ \bibinfo {author} {\bibfnamefont {S.}~\bibnamefont {Yamada}},\ }\bibfield  {title} {\bibinfo {title} {{Collisional and fast neutrino flavor instabilities in two-dimensional core-collapse supernova simulation with Boltzmann neutrino transport}},\ }\href {https://doi.org/10.1103/PhysRevD.109.023012} {\bibfield  {journal} {\bibinfo  {journal} {Phys. Rev. D}\ }\textbf {\bibinfo {volume} {109}},\ \bibinfo {pages} {023012} (\bibinfo {year} {2024})},\ \Eprint {https://arxiv.org/abs/2311.11272} {arXiv:2311.11272 [astro-ph.HE]} \BibitemShut {NoStop}%
\bibitem [{\citenamefont {Liu}\ \emph {et~al.}(2024)\citenamefont {Liu}, \citenamefont {Nagakura}, \citenamefont {Akaho}, \citenamefont {Ito}, \citenamefont {Zaizen}, \citenamefont {Furusawa},\ and\ \citenamefont {Yamada}}]{Liu:2024wzd}%
  \BibitemOpen
  \bibfield  {author} {\bibinfo {author} {\bibfnamefont {J.}~\bibnamefont {Liu}}, \bibinfo {author} {\bibfnamefont {H.}~\bibnamefont {Nagakura}}, \bibinfo {author} {\bibfnamefont {R.}~\bibnamefont {Akaho}}, \bibinfo {author} {\bibfnamefont {A.}~\bibnamefont {Ito}}, \bibinfo {author} {\bibfnamefont {M.}~\bibnamefont {Zaizen}}, \bibinfo {author} {\bibfnamefont {S.}~\bibnamefont {Furusawa}},\ and\ \bibinfo {author} {\bibfnamefont {S.}~\bibnamefont {Yamada}},\ }\bibfield  {title} {\bibinfo {title} {{Muon-induced collisional flavor instability in core-collapse supernova}},\ }\href {https://doi.org/10.1103/PhysRevD.110.043039} {\bibfield  {journal} {\bibinfo  {journal} {Phys. Rev. D}\ }\textbf {\bibinfo {volume} {110}},\ \bibinfo {pages} {043039} (\bibinfo {year} {2024})},\ \Eprint {https://arxiv.org/abs/2407.10604} {arXiv:2407.10604 [hep-ph]} \BibitemShut {NoStop}%
\bibitem [{\citenamefont {Xiong}\ \emph {et~al.}(2024)\citenamefont {Xiong}, \citenamefont {Wu}, \citenamefont {George}, \citenamefont {Lin}, \citenamefont {Largani}, \citenamefont {Fischer},\ and\ \citenamefont {Martinez-Pinedo}}]{Xiong:2024tac}%
  \BibitemOpen
  \bibfield  {author} {\bibinfo {author} {\bibfnamefont {Z.}~\bibnamefont {Xiong}}, \bibinfo {author} {\bibfnamefont {M.-R.}\ \bibnamefont {Wu}}, \bibinfo {author} {\bibfnamefont {M.}~\bibnamefont {George}}, \bibinfo {author} {\bibfnamefont {C.-Y.}\ \bibnamefont {Lin}}, \bibinfo {author} {\bibfnamefont {N.~K.}\ \bibnamefont {Largani}}, \bibinfo {author} {\bibfnamefont {T.}~\bibnamefont {Fischer}},\ and\ \bibinfo {author} {\bibfnamefont {G.}~\bibnamefont {Martinez-Pinedo}},\ }\bibfield  {title} {\bibinfo {title} {{Fast neutrino flavor conversions in a supernova: Emergence, evolution, and effects}},\ }\href {https://doi.org/10.1103/PhysRevD.109.123008} {\bibfield  {journal} {\bibinfo  {journal} {Phys. Rev. D}\ }\textbf {\bibinfo {volume} {109}},\ \bibinfo {pages} {123008} (\bibinfo {year} {2024})},\ \Eprint {https://arxiv.org/abs/2402.19252} {arXiv:2402.19252 [astro-ph.HE]} \BibitemShut {NoStop}%
\bibitem [{\citenamefont {Nagakura}\ \emph {et~al.}(2025)\citenamefont {Nagakura}, \citenamefont {Sumiyoshi}, \citenamefont {Fujibayashi}, \citenamefont {Sekiguchi},\ and\ \citenamefont {Shibata}}]{Nagakura:2025hss}%
  \BibitemOpen
  \bibfield  {author} {\bibinfo {author} {\bibfnamefont {H.}~\bibnamefont {Nagakura}}, \bibinfo {author} {\bibfnamefont {K.}~\bibnamefont {Sumiyoshi}}, \bibinfo {author} {\bibfnamefont {S.}~\bibnamefont {Fujibayashi}}, \bibinfo {author} {\bibfnamefont {Y.}~\bibnamefont {Sekiguchi}},\ and\ \bibinfo {author} {\bibfnamefont {M.}~\bibnamefont {Shibata}},\ }\bibfield  {title} {\bibinfo {title} {{Neutrino flavor instabilities in a binary neutron star merger remnant: Roles of a long-lived hypermassive neutron star}},\ }\Eprint {https://arxiv.org/abs/2504.20143} {arXiv:2504.20143 [astro-ph.HE]}  (\bibinfo {year} {2025})\BibitemShut {NoStop}%
\bibitem [{\citenamefont {Fiorillo}\ \emph {et~al.}(2024)\citenamefont {Fiorillo}, \citenamefont {Padilla-Gay},\ and\ \citenamefont {Raffelt}}]{Fiorillo:2023ajs}%
  \BibitemOpen
  \bibfield  {author} {\bibinfo {author} {\bibfnamefont {D.~F.~G.}\ \bibnamefont {Fiorillo}}, \bibinfo {author} {\bibfnamefont {I.}~\bibnamefont {Padilla-Gay}},\ and\ \bibinfo {author} {\bibfnamefont {G.~G.}\ \bibnamefont {Raffelt}},\ }\bibfield  {title} {\bibinfo {title} {{Collisions and collective flavor conversion: Integrating out the fast dynamics}},\ }\href {https://doi.org/10.1103/PhysRevD.109.063021} {\bibfield  {journal} {\bibinfo  {journal} {Phys. Rev. D}\ }\textbf {\bibinfo {volume} {109}},\ \bibinfo {pages} {063021} (\bibinfo {year} {2024})},\ \Eprint {https://arxiv.org/abs/2312.07612} {arXiv:2312.07612 [hep-ph]} \BibitemShut {NoStop}%
\bibitem [{\citenamefont {Johns}\ and\ \citenamefont {Rodriguez}(2023)}]{Johns:2023xae}%
  \BibitemOpen
  \bibfield  {author} {\bibinfo {author} {\bibfnamefont {L.}~\bibnamefont {Johns}}\ and\ \bibinfo {author} {\bibfnamefont {S.}~\bibnamefont {Rodriguez}},\ }\bibfield  {title} {\bibinfo {title} {{Collisional flavor pendula and neutrino quantum thermodynamics}},\ }\Eprint {https://arxiv.org/abs/2312.10340} {arXiv:2312.10340 [hep-ph]}  (\bibinfo {year} {2023})\BibitemShut {NoStop}%
\bibitem [{\citenamefont {Zaizen}(2025)}]{Zaizen:2025ptx}%
  \BibitemOpen
  \bibfield  {author} {\bibinfo {author} {\bibfnamefont {M.}~\bibnamefont {Zaizen}},\ }\bibfield  {title} {\bibinfo {title} {{Spectral diversity in collisional neutrino-flavor conversion: Flavor equipartition or swap}},\ }\href {https://doi.org/10.1103/PhysRevD.111.103029} {\bibfield  {journal} {\bibinfo  {journal} {Phys. Rev. D}\ }\textbf {\bibinfo {volume} {111}},\ \bibinfo {pages} {103029} (\bibinfo {year} {2025})},\ \Eprint {https://arxiv.org/abs/2502.09260} {arXiv:2502.09260 [hep-ph]} \BibitemShut {NoStop}%
\bibitem [{\citenamefont {Kato}\ \emph {et~al.}(2024)\citenamefont {Kato}, \citenamefont {Nagakura},\ and\ \citenamefont {Johns}}]{Kato:2023cig}%
  \BibitemOpen
  \bibfield  {author} {\bibinfo {author} {\bibfnamefont {C.}~\bibnamefont {Kato}}, \bibinfo {author} {\bibfnamefont {H.}~\bibnamefont {Nagakura}},\ and\ \bibinfo {author} {\bibfnamefont {L.}~\bibnamefont {Johns}},\ }\bibfield  {title} {\bibinfo {title} {{Collisional flavor swap with neutrino self-interactions}},\ }\href {https://doi.org/10.1103/PhysRevD.109.103009} {\bibfield  {journal} {\bibinfo  {journal} {Phys. Rev. D}\ }\textbf {\bibinfo {volume} {109}},\ \bibinfo {pages} {103009} (\bibinfo {year} {2024})},\ \Eprint {https://arxiv.org/abs/2309.02619} {arXiv:2309.02619 [astro-ph.HE]} \BibitemShut {NoStop}%
\bibitem [{\citenamefont {Sigl}\ and\ \citenamefont {Raffelt}(1993)}]{Sigl:1993ctk}%
  \BibitemOpen
  \bibfield  {author} {\bibinfo {author} {\bibfnamefont {G.}~\bibnamefont {Sigl}}\ and\ \bibinfo {author} {\bibfnamefont {G.}~\bibnamefont {Raffelt}},\ }\bibfield  {title} {\bibinfo {title} {{General kinetic description of relativistic mixed neutrinos}},\ }\href {https://doi.org/10.1016/0550-3213(93)90175-O} {\bibfield  {journal} {\bibinfo  {journal} {Nucl. Phys. B}\ }\textbf {\bibinfo {volume} {406}},\ \bibinfo {pages} {423} (\bibinfo {year} {1993})}\BibitemShut {NoStop}%
\bibitem [{\citenamefont {Richers}\ \emph {et~al.}(2019)\citenamefont {Richers}, \citenamefont {McLaughlin}, \citenamefont {Kneller},\ and\ \citenamefont {Vlasenko}}]{Richers:2019grc}%
  \BibitemOpen
  \bibfield  {author} {\bibinfo {author} {\bibfnamefont {S.~A.}\ \bibnamefont {Richers}}, \bibinfo {author} {\bibfnamefont {G.~C.}\ \bibnamefont {McLaughlin}}, \bibinfo {author} {\bibfnamefont {J.~P.}\ \bibnamefont {Kneller}},\ and\ \bibinfo {author} {\bibfnamefont {A.}~\bibnamefont {Vlasenko}},\ }\bibfield  {title} {\bibinfo {title} {{Neutrino Quantum Kinetics in Compact Objects}},\ }\href {https://doi.org/10.1103/PhysRevD.99.123014} {\bibfield  {journal} {\bibinfo  {journal} {Phys. Rev. D}\ }\textbf {\bibinfo {volume} {99}},\ \bibinfo {pages} {123014} (\bibinfo {year} {2019})},\ \bibinfo {note} {[Erratum: \href{https://doi.org/10.1103/PhysRevD.109.129902}{Phys. Rev. D \textbf{109}, 129902(E) (2024)}]},\ \Eprint {https://arxiv.org/abs/1903.00022} {arXiv:1903.00022 [astro-ph.HE]} \BibitemShut {NoStop}%
\bibitem [{\citenamefont {Zhang}\ and\ \citenamefont {Burrows}(2013)}]{Zhang:2013lka}%
  \BibitemOpen
  \bibfield  {author} {\bibinfo {author} {\bibfnamefont {Y.}~\bibnamefont {Zhang}}\ and\ \bibinfo {author} {\bibfnamefont {A.}~\bibnamefont {Burrows}},\ }\bibfield  {title} {\bibinfo {title} {{Transport Equations for Oscillating Neutrinos}},\ }\href {https://doi.org/10.1103/PhysRevD.88.105009} {\bibfield  {journal} {\bibinfo  {journal} {Phys. Rev. D}\ }\textbf {\bibinfo {volume} {88}},\ \bibinfo {pages} {105009} (\bibinfo {year} {2013})},\ \Eprint {https://arxiv.org/abs/1310.2164} {arXiv:1310.2164 [hep-ph]} \BibitemShut {NoStop}%
\bibitem [{\citenamefont {Navas}\ \emph {et~al.}(2024)\citenamefont {Navas} \emph {et~al.}}]{ParticleDataGroup:2024cfk}%
  \BibitemOpen
  \bibfield  {author} {\bibinfo {author} {\bibfnamefont {S.}~\bibnamefont {Navas}} \emph {et~al.} (\bibinfo {collaboration} {Particle Data Group}),\ }\bibfield  {title} {\bibinfo {title} {{Review of particle physics}},\ }\href {https://doi.org/10.1103/PhysRevD.110.030001} {\bibfield  {journal} {\bibinfo  {journal} {Phys. Rev. D}\ }\textbf {\bibinfo {volume} {110}},\ \bibinfo {pages} {030001} (\bibinfo {year} {2024})}\BibitemShut {NoStop}%
\bibitem [{\citenamefont {Foucart}\ \emph {et~al.}(2024)\citenamefont {Foucart}, \citenamefont {Cheong}, \citenamefont {Duez}, \citenamefont {Kidder}, \citenamefont {Pfeiffer},\ and\ \citenamefont {Scheel}}]{Foucart:2024npn}%
  \BibitemOpen
  \bibfield  {author} {\bibinfo {author} {\bibfnamefont {F.}~\bibnamefont {Foucart}}, \bibinfo {author} {\bibfnamefont {P.~C.-K.}\ \bibnamefont {Cheong}}, \bibinfo {author} {\bibfnamefont {M.~D.}\ \bibnamefont {Duez}}, \bibinfo {author} {\bibfnamefont {L.~E.}\ \bibnamefont {Kidder}}, \bibinfo {author} {\bibfnamefont {H.~P.}\ \bibnamefont {Pfeiffer}},\ and\ \bibinfo {author} {\bibfnamefont {M.~A.}\ \bibnamefont {Scheel}},\ }\bibfield  {title} {\bibinfo {title} {{Robustness of neutron star merger simulations to changes in neutrino transport and neutrino-matter interactions}},\ }\href {https://doi.org/10.1103/PhysRevD.110.083028} {\bibfield  {journal} {\bibinfo  {journal} {Phys. Rev. D}\ }\textbf {\bibinfo {volume} {110}},\ \bibinfo {pages} {083028} (\bibinfo {year} {2024})},\ \Eprint {https://arxiv.org/abs/2407.15989} {arXiv:2407.15989 [astro-ph.HE]} \BibitemShut {NoStop}%
\bibitem [{\citenamefont {O'Connor}(2015)}]{OConnor:2014sgn}%
  \BibitemOpen
  \bibfield  {author} {\bibinfo {author} {\bibfnamefont {E.}~\bibnamefont {O'Connor}},\ }\bibfield  {title} {\bibinfo {title} {{An Open-Source Neutrino Radiation Hydrodynamics Code for Core-Collapse Supernovae}},\ }\href {https://doi.org/10.1088/0067-0049/219/2/24} {\bibfield  {journal} {\bibinfo  {journal} {Astrophys. J. Suppl.}\ }\textbf {\bibinfo {volume} {219}},\ \bibinfo {pages} {24} (\bibinfo {year} {2015})},\ \Eprint {https://arxiv.org/abs/1411.7058} {arXiv:1411.7058 [astro-ph.HE]} \BibitemShut {NoStop}%
\bibitem [{\citenamefont {Virtanen}\ \emph {et~al.}(2020)\citenamefont {Virtanen} \emph {et~al.}}]{2020SciPy-NMeth}%
  \BibitemOpen
  \bibfield  {author} {\bibinfo {author} {\bibfnamefont {P.}~\bibnamefont {Virtanen}} \emph {et~al.},\ }\bibfield  {title} {\bibinfo {title} {{{SciPy} 1.0: Fundamental Algorithms for Scientific Computing in Python}},\ }\href {https://doi.org/10.1038/s41592-019-0686-2} {\bibfield  {journal} {\bibinfo  {journal} {Nature Methods}\ }\textbf {\bibinfo {volume} {17}},\ \bibinfo {pages} {261} (\bibinfo {year} {2020})}\BibitemShut {NoStop}%
\bibitem [{\citenamefont {Fiorillo}\ and\ \citenamefont {Raffelt}(2024{\natexlab{a}})}]{Fiorillo:2024qbl}%
  \BibitemOpen
  \bibfield  {author} {\bibinfo {author} {\bibfnamefont {D.~F.~G.}\ \bibnamefont {Fiorillo}}\ and\ \bibinfo {author} {\bibfnamefont {G.~G.}\ \bibnamefont {Raffelt}},\ }\bibfield  {title} {\bibinfo {title} {{Fast Flavor Conversions at the Edge of Instability in a Two-Beam Model}},\ }\href {https://doi.org/10.1103/PhysRevLett.133.221004} {\bibfield  {journal} {\bibinfo  {journal} {Phys. Rev. Lett.}\ }\textbf {\bibinfo {volume} {133}},\ \bibinfo {pages} {221004} (\bibinfo {year} {2024}{\natexlab{a}})},\ \Eprint {https://arxiv.org/abs/2403.12189} {arXiv:2403.12189 [hep-ph]} \BibitemShut {NoStop}%
\bibitem [{\citenamefont {Fiorillo}\ and\ \citenamefont {Raffelt}(2024{\natexlab{b}})}]{Fiorillo:2024uki}%
  \BibitemOpen
  \bibfield  {author} {\bibinfo {author} {\bibfnamefont {D.~F.~G.}\ \bibnamefont {Fiorillo}}\ and\ \bibinfo {author} {\bibfnamefont {G.~G.}\ \bibnamefont {Raffelt}},\ }\bibfield  {title} {\bibinfo {title} {{Theory of neutrino fast flavor evolution. Part II. Solutions at the edge of instability}},\ }\href {https://doi.org/10.1007/JHEP12(2024)205} {\bibfield  {journal} {\bibinfo  {journal} {J. High Energy Phys.}\ }\textbf {\bibinfo {volume} {12}},\ \bibinfo {pages} {205}},\ \Eprint {https://arxiv.org/abs/2409.17232} {arXiv:2409.17232 [hep-ph]} \BibitemShut {NoStop}%
\bibitem [{\citenamefont {Fiorillo}\ and\ \citenamefont {Raffelt}(2025{\natexlab{a}})}]{Fiorillo:2024pns}%
  \BibitemOpen
  \bibfield  {author} {\bibinfo {author} {\bibfnamefont {D.~F.~G.}\ \bibnamefont {Fiorillo}}\ and\ \bibinfo {author} {\bibfnamefont {G.~G.}\ \bibnamefont {Raffelt}},\ }\bibfield  {title} {\bibinfo {title} {{Theory of neutrino slow flavor evolution. Part I. Homogeneous medium}},\ }\href {https://doi.org/10.1007/JHEP04(2025)146} {\bibfield  {journal} {\bibinfo  {journal} {JHEP}\ }\textbf {\bibinfo {volume} {04}},\ \bibinfo {pages} {146}},\ \Eprint {https://arxiv.org/abs/2412.02747} {arXiv:2412.02747 [hep-ph]} \BibitemShut {NoStop}%
\bibitem [{\citenamefont {Fiorillo}\ and\ \citenamefont {Raffelt}(2025{\natexlab{b}})}]{Fiorillo:2025npi}%
  \BibitemOpen
  \bibfield  {author} {\bibinfo {author} {\bibfnamefont {D.~F.~G.}\ \bibnamefont {Fiorillo}}\ and\ \bibinfo {author} {\bibfnamefont {G.~G.}\ \bibnamefont {Raffelt}},\ }\bibfield  {title} {\bibinfo {title} {{Collective Flavor Conversions Are Interactions of Neutrinos with Quantized Flavor Waves}},\ }\href {https://doi.org/10.1103/PhysRevLett.134.211003} {\bibfield  {journal} {\bibinfo  {journal} {Phys. Rev. Lett.}\ }\textbf {\bibinfo {volume} {134}},\ \bibinfo {pages} {211003} (\bibinfo {year} {2025}{\natexlab{b}})},\ \Eprint {https://arxiv.org/abs/2502.06935} {arXiv:2502.06935 [hep-ph]} \BibitemShut {NoStop}%
\bibitem [{\citenamefont {Johns}(2023{\natexlab{b}})}]{Johns:2023jjt}%
  \BibitemOpen
  \bibfield  {author} {\bibinfo {author} {\bibfnamefont {L.}~\bibnamefont {Johns}},\ }\bibfield  {title} {\bibinfo {title} {{Thermodynamics of oscillating neutrinos}},\ }\Eprint {https://arxiv.org/abs/2306.14982} {arXiv:2306.14982 [hep-ph]}  (\bibinfo {year} {2023}{\natexlab{b}})\BibitemShut {NoStop}%
\bibitem [{\citenamefont {Bollig}\ \emph {et~al.}(2017)\citenamefont {Bollig}, \citenamefont {Janka}, \citenamefont {Lohs}, \citenamefont {Martinez-Pinedo}, \citenamefont {Horowitz},\ and\ \citenamefont {Melson}}]{Bollig:2017lki}%
  \BibitemOpen
  \bibfield  {author} {\bibinfo {author} {\bibfnamefont {R.}~\bibnamefont {Bollig}}, \bibinfo {author} {\bibfnamefont {H.~T.}\ \bibnamefont {Janka}}, \bibinfo {author} {\bibfnamefont {A.}~\bibnamefont {Lohs}}, \bibinfo {author} {\bibfnamefont {G.}~\bibnamefont {Martinez-Pinedo}}, \bibinfo {author} {\bibfnamefont {C.~J.}\ \bibnamefont {Horowitz}},\ and\ \bibinfo {author} {\bibfnamefont {T.}~\bibnamefont {Melson}},\ }\bibfield  {title} {\bibinfo {title} {{Muon Creation in Supernova Matter Facilitates Neutrino-driven Explosions}},\ }\href {https://doi.org/10.1103/PhysRevLett.119.242702} {\bibfield  {journal} {\bibinfo  {journal} {Phys. Rev. Lett.}\ }\textbf {\bibinfo {volume} {119}},\ \bibinfo {pages} {242702} (\bibinfo {year} {2017})},\ \Eprint {https://arxiv.org/abs/1706.04630} {arXiv:1706.04630 [astro-ph.HE]} \BibitemShut {NoStop}%
\bibitem [{\citenamefont {Fischer}\ \emph {et~al.}(2020)\citenamefont {Fischer}, \citenamefont {Guo}, \citenamefont {Martinez-Pinedo}, \citenamefont {Liebendörfer},\ and\ \citenamefont {Mezzacappa}}]{Fischer:2020vie}%
  \BibitemOpen
  \bibfield  {author} {\bibinfo {author} {\bibfnamefont {T.}~\bibnamefont {Fischer}}, \bibinfo {author} {\bibfnamefont {G.}~\bibnamefont {Guo}}, \bibinfo {author} {\bibfnamefont {G.}~\bibnamefont {Martinez-Pinedo}}, \bibinfo {author} {\bibfnamefont {M.}~\bibnamefont {Liebendörfer}},\ and\ \bibinfo {author} {\bibfnamefont {A.}~\bibnamefont {Mezzacappa}},\ }\bibfield  {title} {\bibinfo {title} {{Muonization of supernova matter}},\ }\href {https://doi.org/10.1103/PhysRevD.102.123001} {\bibfield  {journal} {\bibinfo  {journal} {Phys. Rev. D}\ }\textbf {\bibinfo {volume} {102}},\ \bibinfo {pages} {123001} (\bibinfo {year} {2020})},\ \Eprint {https://arxiv.org/abs/2008.13628} {arXiv:2008.13628 [astro-ph.HE]} \BibitemShut {NoStop}%
\bibitem [{\citenamefont {Loffredo}\ \emph {et~al.}(2023)\citenamefont {Loffredo}, \citenamefont {Perego}, \citenamefont {Logoteta},\ and\ \citenamefont {Branchesi}}]{Loffredo:2022prq}%
  \BibitemOpen
  \bibfield  {author} {\bibinfo {author} {\bibfnamefont {E.}~\bibnamefont {Loffredo}}, \bibinfo {author} {\bibfnamefont {A.}~\bibnamefont {Perego}}, \bibinfo {author} {\bibfnamefont {D.}~\bibnamefont {Logoteta}},\ and\ \bibinfo {author} {\bibfnamefont {M.}~\bibnamefont {Branchesi}},\ }\bibfield  {title} {\bibinfo {title} {{Muons in the aftermath of neutron star mergers and their impact on trapped neutrinos}},\ }\href {https://doi.org/10.1051/0004-6361/202244927} {\bibfield  {journal} {\bibinfo  {journal} {Astron. Astrophys.}\ }\textbf {\bibinfo {volume} {672}},\ \bibinfo {pages} {A124} (\bibinfo {year} {2023})},\ \Eprint {https://arxiv.org/abs/2209.04458} {arXiv:2209.04458 [astro-ph.HE]} \BibitemShut {NoStop}%
\bibitem [{\citenamefont {Ng}\ \emph {et~al.}(2025)\citenamefont {Ng}, \citenamefont {Musolino}, \citenamefont {Tootle},\ and\ \citenamefont {Rezzolla}}]{Ng:2024zve}%
  \BibitemOpen
  \bibfield  {author} {\bibinfo {author} {\bibfnamefont {H.~H.-Y.}\ \bibnamefont {Ng}}, \bibinfo {author} {\bibfnamefont {C.}~\bibnamefont {Musolino}}, \bibinfo {author} {\bibfnamefont {S.~D.}\ \bibnamefont {Tootle}},\ and\ \bibinfo {author} {\bibfnamefont {L.}~\bibnamefont {Rezzolla}},\ }\bibfield  {title} {\bibinfo {title} {{Accurate Muonic Interactions in Neutron Star Mergers and Impact on Heavy-element Nucleosynthesis}},\ }\href {https://doi.org/10.3847/2041-8213/add324} {\bibfield  {journal} {\bibinfo  {journal} {Astrophys. J. Lett.}\ }\textbf {\bibinfo {volume} {985}},\ \bibinfo {pages} {L36} (\bibinfo {year} {2025})},\ \Eprint {https://arxiv.org/abs/2411.19178} {arXiv:2411.19178 [astro-ph.HE]} \BibitemShut {NoStop}%
\bibitem [{\citenamefont {Pajkos}\ and\ \citenamefont {Most}(2025)}]{Pajkos:2025oyf}%
  \BibitemOpen
  \bibfield  {author} {\bibinfo {author} {\bibfnamefont {M.~A.}\ \bibnamefont {Pajkos}}\ and\ \bibinfo {author} {\bibfnamefont {E.~R.}\ \bibnamefont {Most}},\ }\bibfield  {title} {\bibinfo {title} {{Influence of muons, pions, and trapped neutrinos on neutron star mergers}},\ }\href {https://doi.org/10.1103/PhysRevD.111.043013} {\bibfield  {journal} {\bibinfo  {journal} {Phys. Rev. D}\ }\textbf {\bibinfo {volume} {111}},\ \bibinfo {pages} {043013} (\bibinfo {year} {2025})},\ \Eprint {https://arxiv.org/abs/2409.09147} {arXiv:2409.09147 [astro-ph.HE]} \BibitemShut {NoStop}%
\bibitem [{\citenamefont {Liu}\ \emph {et~al.}(2025)\citenamefont {Liu}, \citenamefont {Nagakura}, \citenamefont {Zaizen}, \citenamefont {Johns},\ and\ \citenamefont {Yamada}}]{Liu:2025tnf}%
  \BibitemOpen
  \bibfield  {author} {\bibinfo {author} {\bibfnamefont {J.}~\bibnamefont {Liu}}, \bibinfo {author} {\bibfnamefont {H.}~\bibnamefont {Nagakura}}, \bibinfo {author} {\bibfnamefont {M.}~\bibnamefont {Zaizen}}, \bibinfo {author} {\bibfnamefont {L.}~\bibnamefont {Johns}},\ and\ \bibinfo {author} {\bibfnamefont {S.}~\bibnamefont {Yamada}},\ }\bibfield  {title} {\bibinfo {title} {{Asymptotic states of fast neutrino-flavor conversions in the three-flavor framework}},\ }\href {https://doi.org/10.1103/v9lr-ydbb} {\bibfield  {journal} {\bibinfo  {journal} {Phys. Rev. D}\ }\textbf {\bibinfo {volume} {111}},\ \bibinfo {pages} {123004} (\bibinfo {year} {2025})},\ \Eprint {https://arxiv.org/abs/2503.18145} {arXiv:2503.18145 [astro-ph.HE]} \BibitemShut {NoStop}%
\bibitem [{\citenamefont {Bhattacharyya}\ \emph {et~al.}(2025)\citenamefont {Bhattacharyya}, \citenamefont {Wu},\ and\ \citenamefont {Xiong}}]{Bhattacharyya:2025gds}%
  \BibitemOpen
  \bibfield  {author} {\bibinfo {author} {\bibfnamefont {S.}~\bibnamefont {Bhattacharyya}}, \bibinfo {author} {\bibfnamefont {M.-R.}\ \bibnamefont {Wu}},\ and\ \bibinfo {author} {\bibfnamefont {Z.}~\bibnamefont {Xiong}},\ }\bibfield  {title} {\bibinfo {title} {{Role of Matter Inhomogeneity on Fast Flavor Conversion of Supernova Neutrinos}},\ }\Eprint {https://arxiv.org/abs/2504.11316} {arXiv:2504.11316 [astro-ph.HE]}  (\bibinfo {year} {2025})\BibitemShut {NoStop}%
\bibitem [{\citenamefont {Froustey}\ \emph {et~al.}(2020)\citenamefont {Froustey}, \citenamefont {Pitrou},\ and\ \citenamefont {Volpe}}]{Froustey:2020mcq}%
  \BibitemOpen
  \bibfield  {author} {\bibinfo {author} {\bibfnamefont {J.}~\bibnamefont {Froustey}}, \bibinfo {author} {\bibfnamefont {C.}~\bibnamefont {Pitrou}},\ and\ \bibinfo {author} {\bibfnamefont {M.~C.}\ \bibnamefont {Volpe}},\ }\bibfield  {title} {\bibinfo {title} {{Neutrino decoupling including flavour oscillations and primordial nucleosynthesis}},\ }\href {https://doi.org/10.1088/1475-7516/2020/12/015} {\bibfield  {journal} {\bibinfo  {journal} {JCAP}\ }\textbf {\bibinfo {volume} {12}},\ \bibinfo {pages} {015}},\ \Eprint {https://arxiv.org/abs/2008.01074} {arXiv:2008.01074 [hep-ph]} \BibitemShut {NoStop}%
\bibitem [{\citenamefont {Froustey}\ and\ \citenamefont {Pitrou}(2022)}]{Froustey:2021azz}%
  \BibitemOpen
  \bibfield  {author} {\bibinfo {author} {\bibfnamefont {J.}~\bibnamefont {Froustey}}\ and\ \bibinfo {author} {\bibfnamefont {C.}~\bibnamefont {Pitrou}},\ }\bibfield  {title} {\bibinfo {title} {{Primordial neutrino asymmetry evolution with full mean-field effects and collisions}},\ }\href {https://doi.org/10.1088/1475-7516/2022/03/065} {\bibfield  {journal} {\bibinfo  {journal} {JCAP}\ }\textbf {\bibinfo {volume} {03}}\bibfield  {number} {\bibinfo  {number} { (03)},\ \bibinfo {pages} {065}},\ }\Eprint {https://arxiv.org/abs/2110.11889} {arXiv:2110.11889 [hep-ph]} \BibitemShut {NoStop}%
\end{thebibliography}%

\end{document}